\documentclass[11pat, letterpaper]{article}
\usepackage[margin=1in]{geometry}
\usepackage[utf8]{inputenc} %

\usepackage{booktabs} %
\usepackage[table,xcdraw]{xcolor}
\usepackage{bbm}

\usepackage{forloop}
\usepackage[ruled, vlined]{algorithm2e} %
\usepackage{algpseudocode}

\usepackage{mathtools}
\usepackage{macros_ms} %
\usepackage{fullpage}

\definecolor{skyblue}{RGB}{135,206,235} %

\usepackage[labelfont=bf,font={footnotesize,stretch=0.90}]{caption}
\usepackage[labelformat=simple]{subcaption}

\usepackage{ragged2e}
\usepackage[round]{natbib}
\usepackage[english]{babel}
\usepackage{amsfonts}
\usepackage{amsmath, amsthm, amssymb}
\usepackage{ifthen}
\usepackage{url}
\usepackage{graphicx}
\usepackage{color}
\usepackage{array}
\usepackage{enumitem}
\usepackage{xspace}
\usepackage{theoremref}

\usepackage{authblk}

\allowdisplaybreaks

\usepackage[
  detect-weight,
  exponent-product=\cdot,
  locale = DE,
  group-separator=.,
]{siunitx}
\DeclareSIUnit{\pp}{\textup{p.p.}}

\usepackage[most]{tcolorbox}

\usepackage[protrusion=true,expansion=true]{}

\usepackage{amsmath}
\usepackage{amsthm}
\usepackage{amssymb}
\usepackage{bbm}
\usepackage{mathtools}
\usepackage{mathrsfs}

\newtheorem{ass}{Assumption}[]

\usepackage{courier} %

\usepackage{lipsum} %

\usepackage{graphicx}
\usepackage{subcaption}
\usepackage[space]{grffile} %

\usepackage{theoremref}
\usepackage{graphicx}
\usepackage{wrapfig}
\usepackage{xcolor}
\definecolor{dark-red}{rgb}{0.4,0.15,0.15}
\definecolor{dark-blue}{rgb}{0,0,0.7}
\hypersetup{
    colorlinks, linkcolor={dark-blue},
    citecolor={dark-blue}, urlcolor={dark-blue}
}

\usepackage{etoc}

\usepackage{algpseudocode}
\SetKwProg{Fn}{Function}{}{}
\let\oldnl\nl%
\newcommand{\nonl}{\renewcommand{\nl}{\let\nl\oldnl}}%

\usepackage{multicol}
\usepackage{enumitem}
\usepackage[utf8]{inputenc} %
\usepackage[T1]{fontenc}    %
\usepackage{hyperref}       %
\usepackage{url}            %
\usepackage{booktabs}       %
\usepackage{nicefrac}       %
\usepackage[parfill]{parskip}

\DeclareMathOperator*{\ind}{\perp\!\!\!\perp}

\DeclareMathOperator*{\E}{\mathbb E}

\DeclarePairedDelimiter\autobracket{(}{)}
\newcommand{\br}[1]{\autobracket*{#1}}

\DeclarePairedDelimiter\autobrackett{[}{]}
\newcommand{\brr}[1]{\autobrackett*{#1}}

\usepackage{authblk}

\title{The GPT Surprise: Offering Large Language Model Chat in a Massive Coding Class Reduced Engagement but Increased Adopters' Exam Performances}

\author[1]{Allen Nie\thanks{Corresponding author emails: \{anie, ebrun, piech\}@cs.stanford.edu}} %
\author[1]{Yash Chandak}
\author[2]{Miroslav Suzara}
\author[1]{Ali Malik}
\author[1]{Juliette Woodrow}
\author[1]{Matt Peng}
\author[1]{\\Mehran Sahami}
\author[1]{Emma Brunskill}
\author[1]{Chris Piech}
\affil[1]{Computer Science, Stanford University}
\affil[2]{Education, Stanford University}
\date{}
\newcommand{\ittEffect}{\textbf{Advertisement Effect}}
\newcommand{\treatEffect}{\textbf{Effect for Adopters}}

\begin{document}

\maketitle

\begin{abstract}
\noindent Large language models (LLMs) are quickly being adopted in a wide range of learning experiences, especially via ubiquitous and broadly accessible chat interfaces like ChatGPT and Copilot. This type of interface is readily available to students and teachers around the world, yet relatively little research has been done to assess the impact of such generic tools on student learning. Coding education is an interesting test case, both because LLMs have strong performance on coding tasks, and because LLM-powered support tools are rapidly becoming part of the workflow of professional software engineers. To help understand the impact of generic LLM use on coding education, we conducted a large-scale randomized control trial with 5,831 students from 146 countries in an online coding class in which we provided some students with access to a chat interface with GPT-4. 
We estimate positive benefits on exam performance for adopters, the students who used the tool, but over all students, the advertisement of GPT-4 led to a significant average decrease in exam participation. We observe similar decreases in other forms of course engagement. However, this decrease is modulated by the student's country of origin. Offering access to LLMs to students from low human development index countries increased their exam participation rate on average.
Our results suggest there may be promising benefits to using LLMs in an introductory coding class, but also potential harms for engagement, which makes their longer term impact on student success unclear.
Our work highlights the need for additional investigations to help understand the potential impact of future adoption and integration of LLMs into classrooms.

\end{abstract}

\section{Introduction}

Large language models (LLMs) have the potential to substantially impact education.
While there are many ways that LLMs could be used to support stakeholders in the educational ecosystem, including training tutors~\citep{markel2023gpteach},  creating content or lesson plans for teachers~\citep{pardos2023learning}, and helping parents better support their kids on their homework~\citep{team2023gemini}, one key opportunity is to support students directly. LLMs can act like tutors, supporting the student in real-time with natural language interactions~\citep{duolingomax2023,khanmigo2023}, which, based on prior evidence~\citep{bloom19842}, could lead to substantially more personalized and effective learning. On the other hand, there is substantial concern that students may use LLMs as a substitute for their own learning, such as by using LLMs to generate homework solutions, though some suggest this concern may so far be overestimated~\citep{stanford2023aichatbots}. Indeed, many students are already using generic (not education-specific) LLMs, such as the publicly available chatbot interfaces provided by large companies, for their own education~\citep{stanforddaily2023chatgpt}. However, we know very little about the impact on student learning when students get support from such a generalist tool at the moment.

In this paper we aim to start answering this question through an experiment in which we provided access to a popular and powerful general LLM, GPT-4, to students in a large online coding course. Coding education is an area where one might expect LLM support to be particularly beneficial for two reasons. First, though LLMs already demonstrate proficiency in a wide range of tasks that can assist a human user, they are known to have strong performance in writing computer programs. Second, it is likely that programmers may be expected to use LLM-based coding support tools, such as Github Copilot\footnote{https://github.com/features/copilot} or JetBrains AI Assistant, in a software engineering job. For these two reasons, one might expect that interacting with a LLM could support a student's learning of both coding concepts, and their ability to leverage LLMs to produce coding. However, though there is evidence that there is a significant productivity boost for the expert programmers~\citep{noy2023experimental}, we know little about how using  LLMs might impact a coding beginner.

To investigate the potential impact of LLMs on students learning to code, we conducted a randomized control trial in which we provided a subset of students in a massive free online coding class with over 8,762 students in 146 countries access to a course-specific interface to GPT-4. 
We provided learners with information about the potential limitations and benefits of using LLMs and designed prompts to prevent students from seeking direct solutions from GPT-4 (Section~\ref{sec:experiment-design}).
To understand how LLMs impacted student learning, we analyzed student engagement in the course, measured by exam participation and validated by homework completion and section attendance, and their performance on the (optional) exam (Section~\ref{sec:low_engagement}). 
The non-compulsory nature of the course made it straightforward to measure changes in engagement, but more complex to understand impacts on student learning.
We used a causal inference estimator to understand the potential learning benefit for the students (Section~\ref{sec:exam_score}). We discuss our findings in Section~\ref{sec:limitations} and \ref{sec:discussion} but highlight the core insights here:

\begin{enumerate}[itemsep=0pt, wide=0pt, leftmargin=*, after=\strut,label={\arabic*.}]
\item \textbf{Low Adoption Rate}: In contrast to concerns that students may overuse large language models, we found only a small number of students (14.2\%) who were provided access to our course GPT interface, and emailed to alert them to this opportunity, chose to use our GPT interface. This might suggest that student adoption and usage of LLMs follow a typical technology adoption curve~\citep{rogers2014diffusion}.
\item \textbf{LLM Lowered Course Engagement} As part of the course, all students were offered an opportunity to take an optional diagnostic exam in a 3-day period. We found that giving students access to and advertising GPT-4 led to a substantial, statistically significant  \underline{decrease} in their exam participation. The effect of access to GPT was negative 4.3 percentage points. Access and advertising also led to a decrease in homework participation and section attendance. This engagement decrease causes concern. However, this trend is not the same across countries. Students' exam participation rate \underline{increased} by 14.8 percentage points when they came from countries with low Human Development Index (HDI) scores (Section~\ref{sec:low_engagement}), suggesting a heterogeneous effect of risks and benefits when we provide LLMs to students.
\item \textbf{LLM Had Positive Estimated Benefits on Learning for Adopters} Low adoption rate overall and the fact that students could freely choose to take the diagnostic exam introduced complexity in understanding the effect of using LLM for learning. We used a causal inference estimator that helped us understand whether there could be a potential positive benefit on learning outcomes, as evaluated by exam scores. Our estimate suggests that by providing GPT-4 to students, the students who are adopters could see a 6.8 average percentage point \underline{increase} in their exam scores over what they would have achieved without using GPT-4 (Table~\ref{table:learning_gain}).
\end{enumerate}

Our findings paint a nuanced picture of how GPT-4 could affect coding education. 
Typically, introducing novel features in a learning setting increases student engagement. Such an effect has been found in many education settings and reproduced in the same course where we conducted our current experiment~\citep{makel2014facts,bigman2021pearprogram,demszky2023can,markel2023gpteach}.
However, offering students LLMs through a chat interface induced the opposite effect.
The significant decrease in student engagement from advertising and providing access to an LLM is unexpected and surprising.
Our result also showed that the effect on engagement is moderated by the student's country of origin. Students from countries with low HDI scores engaged with the class more when they were given access to LLMs. 
This kind of heterogeneous impact of educational intervention has been observed in other learning settings before~\citep{leite2022heterogeneity,nie2023understanding,ruan2024reinforcement}.\
These results suggest that while LLMs have the potential to support some students' learning outcomes, care also must be taken to ensure they do not negatively impact students' willingness to spend more effort in class. 

\begin{figure}
    \centering
    \includegraphics[scale=0.4]{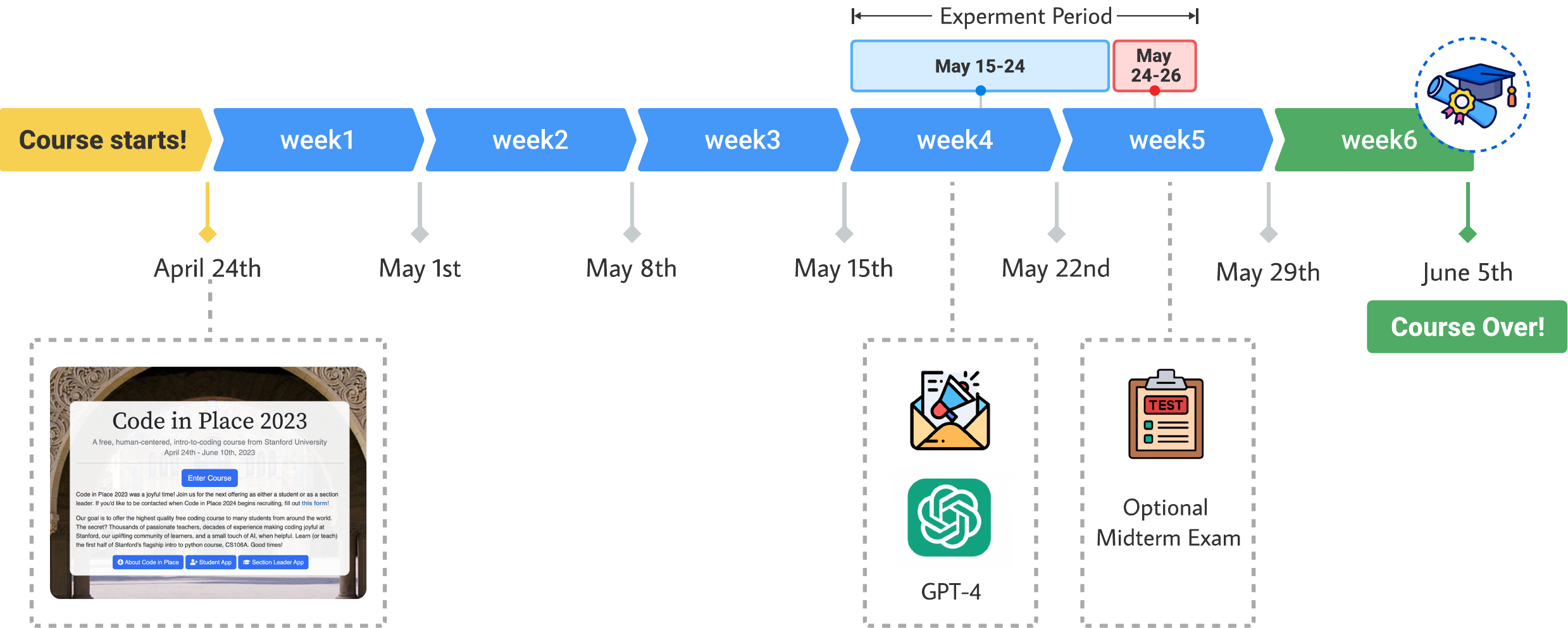}
    \caption{Timeline of the course and the randomized control trial experiment. The course starts on April 24th. Random assignment of students to treatment and control is decided on May 8th. An email is sent to students in the treatment group, letting them know that they now have access to a customized free GPT-4 interface within the class. Students in the control group do not get the email. An optional diagnostic exam is administered between May 24-26.}
    \label{fig:timeline}
\end{figure}

\section{Experiment Design}
\label{sec:experiment-design}

\begin{figure}[t]
    \centering
    \includegraphics[width=0.95\textwidth]{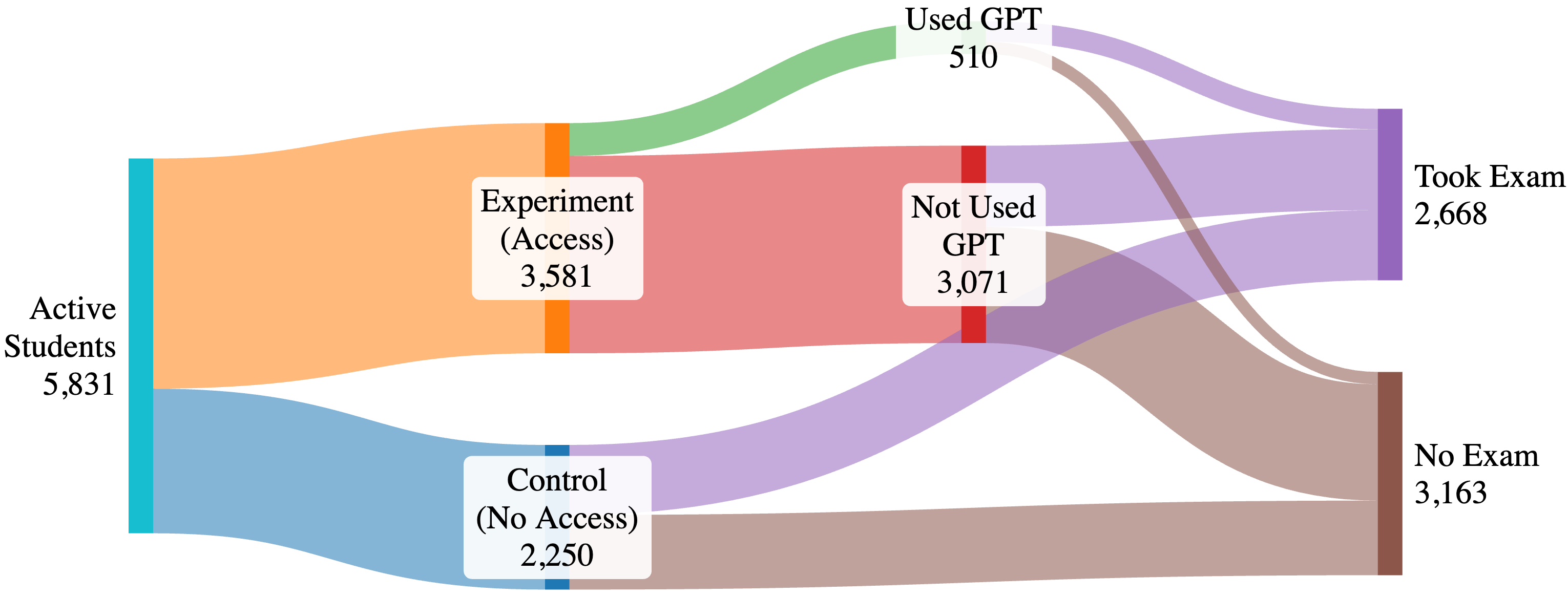}
    \caption{We show the experiment process as a flow diagram. We chose our experiment pool to be students who are still active after week 1 (N=5,831). We randomly assigned 60\% of the students (n=3,581) to the treatment condition and 40\% of the students (n=2,250) to the control condition. Students in the experiment group were given free access to GPT-4 through a customized built-in interface (see Figure~\ref{fig:interface}). Some students took the optional 4-hour midterm exam, and others did not. An after-course survey found around 2\% of students stated they have used ChatGPT outside of the class.}
    \label{fig:population}
\end{figure}

To evaluate the impact of providing GPT-4 to coding beginners taking a free online class, we conducted a randomized control trial study to analyze the impact of GPT-4 by comparing a series of course activities and learning outcomes between the experiment and control group. 

Our experiment consists of 5,831 students (out of 8,762 students who were given approval to enroll in class) who were determined to be active after the first week. We randomly assigned 60\% of the students (n=3,581) to the experiment condition and 40\% of the students (n=2,250) to the control condition. We give slightly more students access to GPT-4 because we expect the measurable educational outcome to have a high variance for the experiment group. We show the demographic information of the experiment and control groups in Table~\ref{tab:demographics}. Students in the experiment group were given access to a customized in-class ChatGPT-like interface with GPT-4 at the start of week 4 in the course (see Figure~\ref{fig:interface}). The students saw an altered home page interface with a ChatGPT button on the sidebar, and we sent an email notification to inform students of its presence (see Figure~\ref{fig:email}). After 9 days (at the end of week 5), all students were invited to take an optional 4-hour midterm exam as part of a regular course offering. Since this exam is not mandatory for a student to earn their course certificate, 45.8\% of students took the exam. Additionally, not every student who was given access to GPT-4 interacted with GPT-4. Only 14.2\% (N=510) students in the experiment group used GPT-4 prior to the midterm exam.
We show this whole process in Figure~\ref{fig:population}.

\begin{table}[ht!]
    \centering
    \small
    \begin{tabular}{lcccc}
        \toprule
        & \begin{tabular}[c]{@{}c@{}}Access to GPT-4 \\ (Experiment) \end{tabular}   & \begin{tabular}[c]{@{}c@{}}No Access to GPT-4 \\ (Control)  \end{tabular} & \begin{tabular}[c]{@{}c@{}} Used GPT-4 \\ (Subset of Experiment) \end{tabular} & 
        \begin{tabular}[c]{@{}c@{}} Not Used GPT-4 \\ (Subset of Experiment) \end{tabular}  \\
        \midrule
        Age &  \begin{tabular}[c]{@{}c@{}}31.4 (10.3) \end{tabular}  & \begin{tabular}[c]{@{}c@{}} 31.4 (10.7)\end{tabular}  & 32.3 (11.6) & 31.2 (10.1)\\ %
        Gender (Female) & 53.9\% & 55.3\% & 48.6\% & 54.7\% \\
        Application Score & 48.2 (11.5) & 48.2 (11.2) & 48.4 (11.6) & 48.2 (11.5)\\
        English Fluency & 13.8 (2.5) & 13.8 (2.5) & 13.7 (2.5) & 13.8 (2.5)\\
        Application Effort & 4 (1.3) & 4 (1.4) & 4.0 (1.3) & 4.0 (1.3) \\
        Coding Score & 8.2 (3.8) & 8.3 (3.8) & 8.4 (3.7) & 8.2 (3.9)\\ 
        Prior Experience & 5.1 (4.4) & 5.2 (4.4) & 5.2 (4.2) & 5.1 (4.4) \\
        Friend Score & 4.7 (18.7) & 4.5 (17.5) & 5.5 (22.6) & 4.6 (18.0) \\
        Section Attendance & 1.65 (0.61) & 1.66 (0.60) & 1.75 (0.53) & 1.64 (0.62) \\
        Country HDI & 0.83 (0.12) & 0.83 (0.12) & 0.81 (0.12) & 0.83 (0.12) \\
        No. Observations & 3581 & 2250 & 510 & 3071 \\
        \midrule
        Exam Participation & 44.1\% & 48.4\% & 62.4\% & 41.1\% \\
        Used In-Class GPT & 14.2\% & 0\% & 100\% & 0\%  \\
        Exam Score & 87.0 (18.7) & 86.1 (20.0) & 88.4 (17.3) & 86.6 (19.0) \\
        \bottomrule
    \end{tabular}
    \caption{We report the mean and standard deviation for the demographic covariate on four groups. The first two groups are students from the experiment and control group. Despite being offered access to GPT-4, some students in the experiment group did not use GPT-4. Therefore, we can have two other groups that are subsets of the experiment group, with students who used and did not use GPT-4. We refer to students in the experiment group who used GPT-4 as ``adopters.''}
    \label{tab:demographics}
\end{table}

\section{Results}
\label{sec:result}

To assess educational impact, we focus on two outcome measures: (1) Student engagement, especially whether or not the student takes the diagnostic exam, and (2) the score they obtained taking the diagnostic exam. The exam consists of 5 multi-part coding problems of increasing difficulty.

There are two types of impact that introducing GPT-4 to students might have. Sending students an email and letting them read through our introduction to GPT-4 might affect a student's behavior even without them directly using the GPT-4 tool. Our introductory text to the students about the tool is detailed and discusses both the positives and negatives of using GPT-4 (Figure~\ref{fig:pos_adv}). We framed GPT-4 as a tool that can make programming easier and offer instant support to their questions but also highlighted the hallucination, the potential harm to their learning, and the possibility of receiving unsettling responses. We refer to this as the \ittEffect{}. However, there is another type of effect, which is whether the act of using GPT-4 impacts student learning. In our paper, we call this the \treatEffect{}. In econometrics, medicine, and statistics, the \ittEffect{} is commonly known as the ``intent-to-treat effect'', and the \treatEffect{} is known as the ``treatment effect on the compliers''. These two effects are the same if every person in the experiment group takes the treatment (i.e., every student who got access to GPT-4 used GPT-4 before the diagnostic exam). We chose this experimental design because we think it is a reasonable model of many settings being considered for LLM to support educational settings, where students will have the option, but not requirement, to use LLM tools (for example, Khan Academy's Khanmigo). 
We define these effects more precisely in Section~\ref{sec:method}.

We first start our analysis by looking at how \textbf{advertising} GPT-4 impacted students on their course participation in Section~\ref{sec:low_engagement}.  We then estimate the intent to treat (\ittEffect{}) on student diagnostic exam scores, and we also estimate the effect on exam scores on those that used our GPT-4 in Section~\ref{sec:exam_score}.

\begin{figure}[ht]
    \centering
    \begin{subfigure}[b]{0.53\linewidth}
        \includegraphics[width=\linewidth]{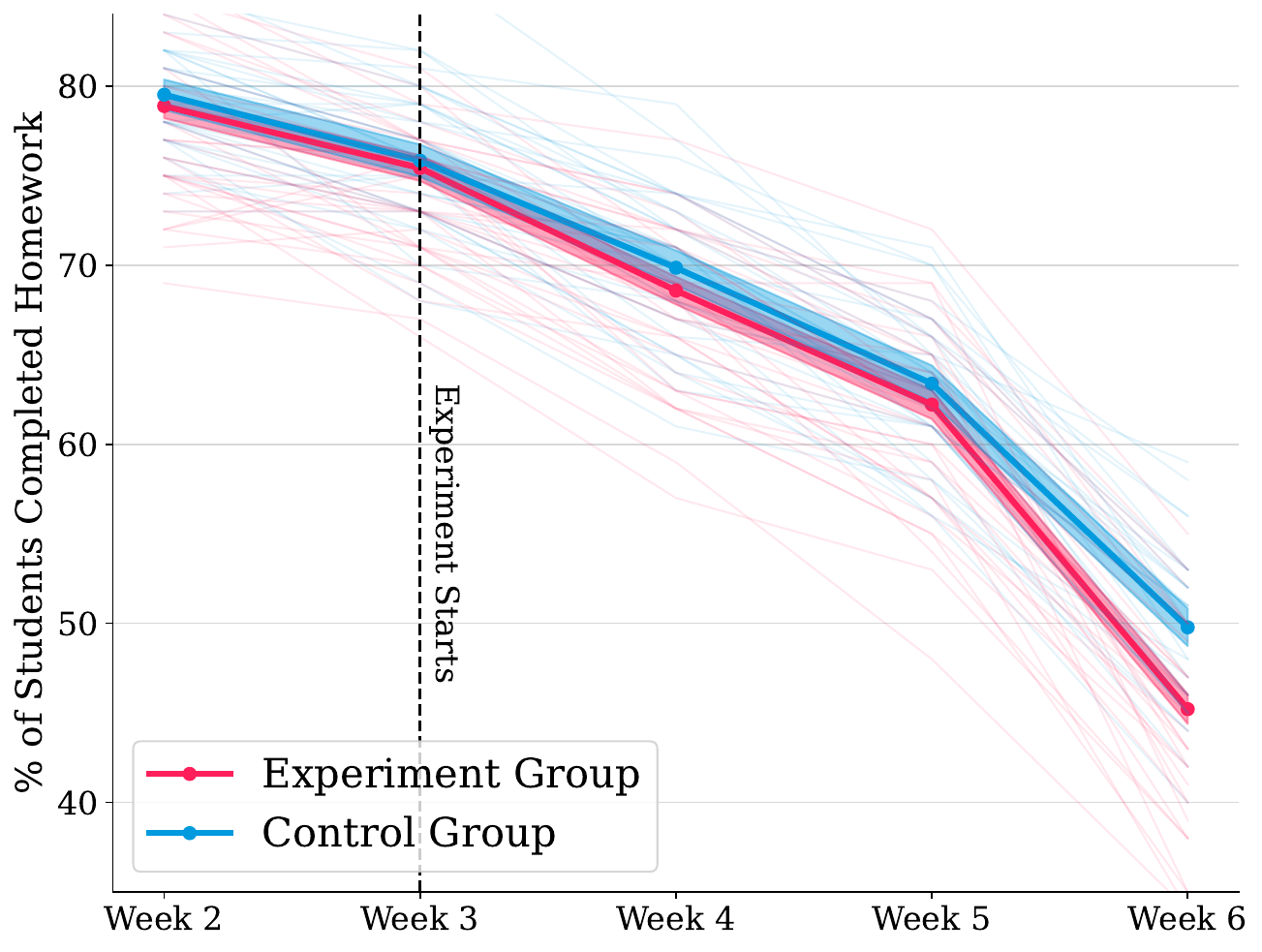}
        \caption{Weekly Homework Completion Drop-off}
        \label{fig:engagement_curve}
    \end{subfigure}
    \begin{subfigure}[b]{0.45\linewidth}
        \includegraphics[width=\linewidth]{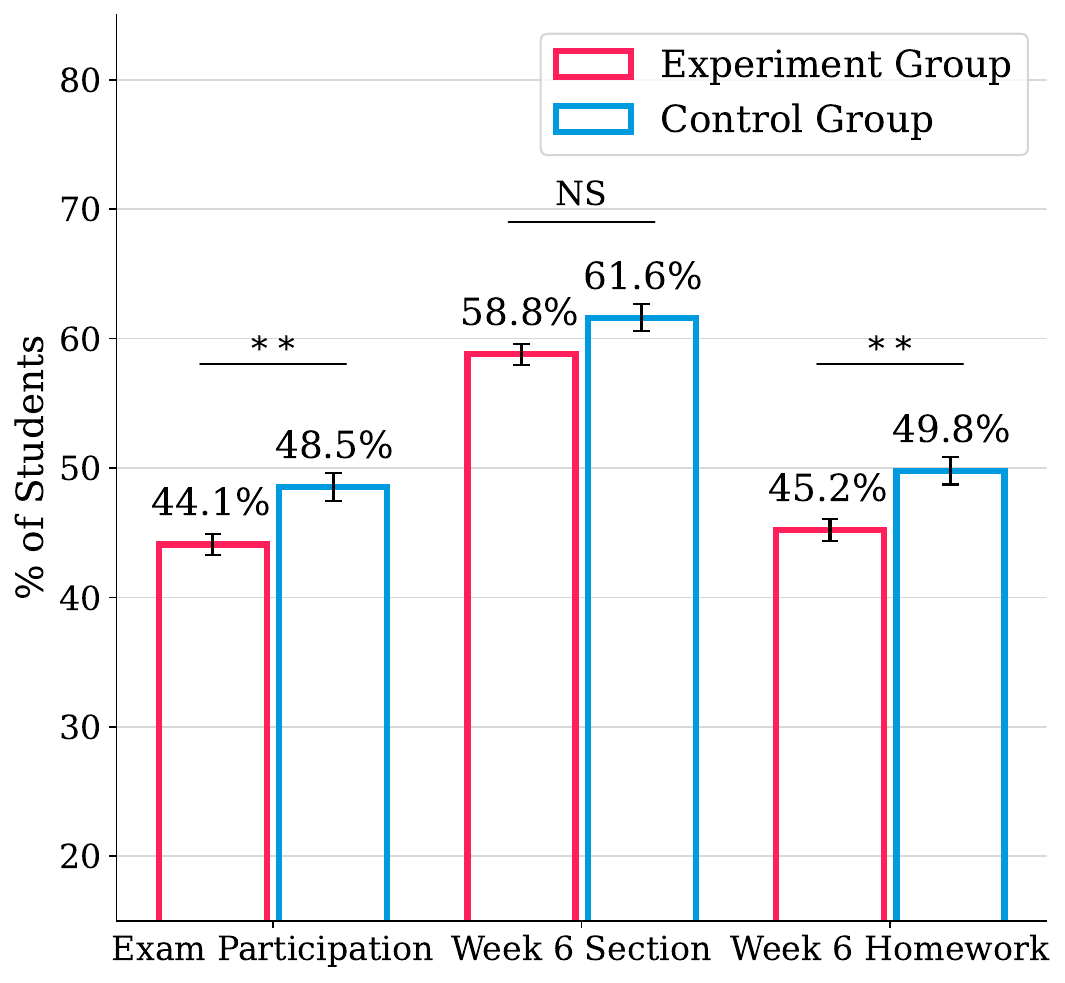}
        \caption{Exam Participation and Week 6 Activities}
        \label{fig:engagement_hist}
    \end{subfigure}
    \caption{(a) We show that the homework completion rate decreases as the course progresses. The shaded area around the mean represents standard error (SE). Light-colored lines are a random subsample of 100 students. Before the experiment started, the two groups had completed homework at a similar rate. After the experiment started, the completion rate began to diverge. (b) A difference between the experiment and control group around exam participation (SE=1.34, CI=(-7.10, -1.82), $P$=0.006), Week 6 section attendance (SE=1.31, CI=(-5.41, -0.26), $P$>0.05), and Week 6 homework completion (SE=1.34, CI=(-7.20, -1.94), $P$=0.005). To account for multiple comparisons, we used Bonferroni correction. ***$P$<0.001, **$P$<0.01, *$P$<0.05, NS $P$>0.05.}
    \label{fig:engagement_main}
\end{figure}

\begin{figure}[ht]
    \centering
    \begin{subfigure}[b]{0.32\linewidth}
        \includegraphics[width=\linewidth]{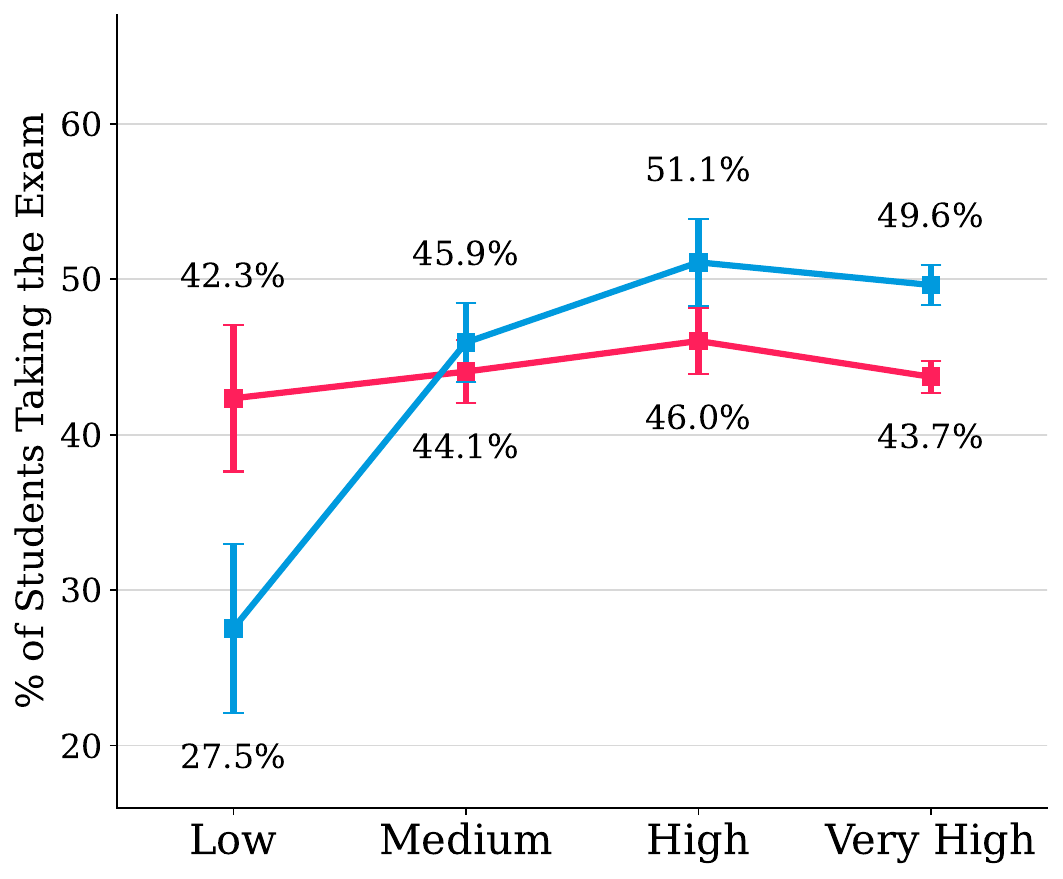}
        \caption{Student Country United Nations HDI}
        \label{fig:exam_hdi}
    \end{subfigure}
    \begin{subfigure}[b]{0.32\linewidth}
        \includegraphics[width=\linewidth]{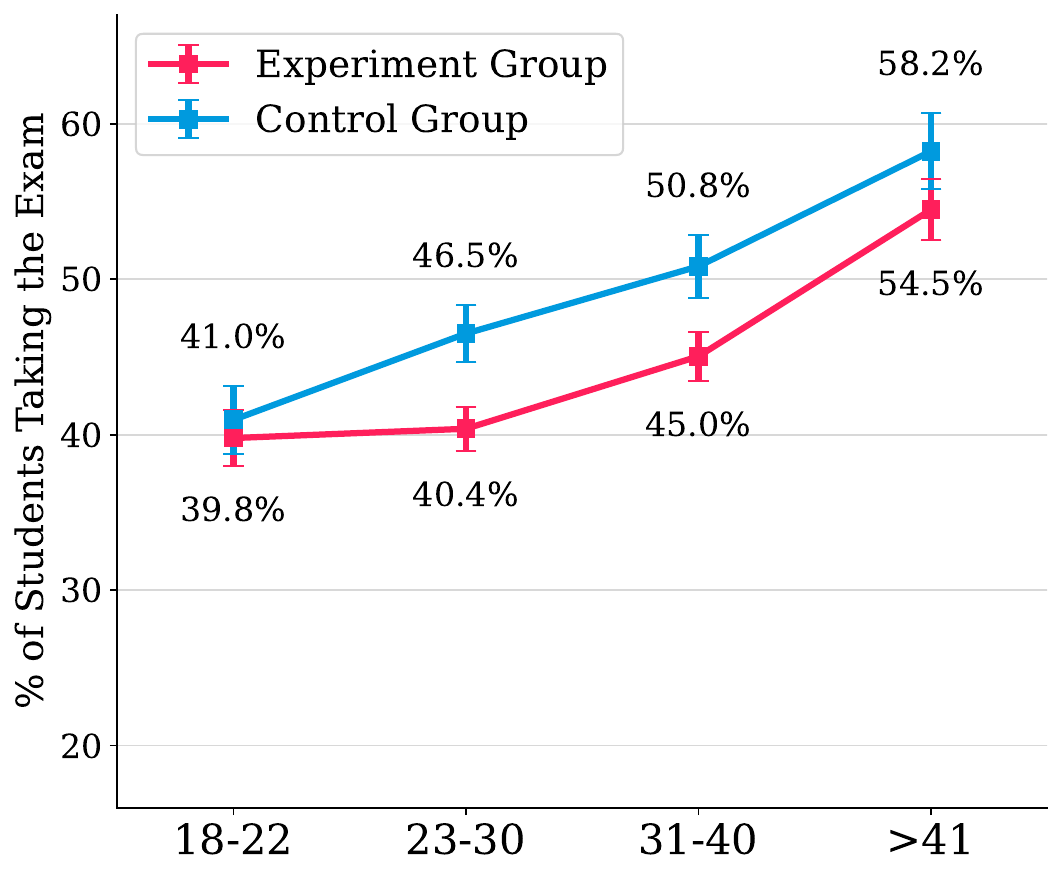}
        \caption{Student Age}
        \label{fig:exam_age}
    \end{subfigure}
    \begin{subfigure}[b]{0.32\linewidth}
        \includegraphics[width=\linewidth]{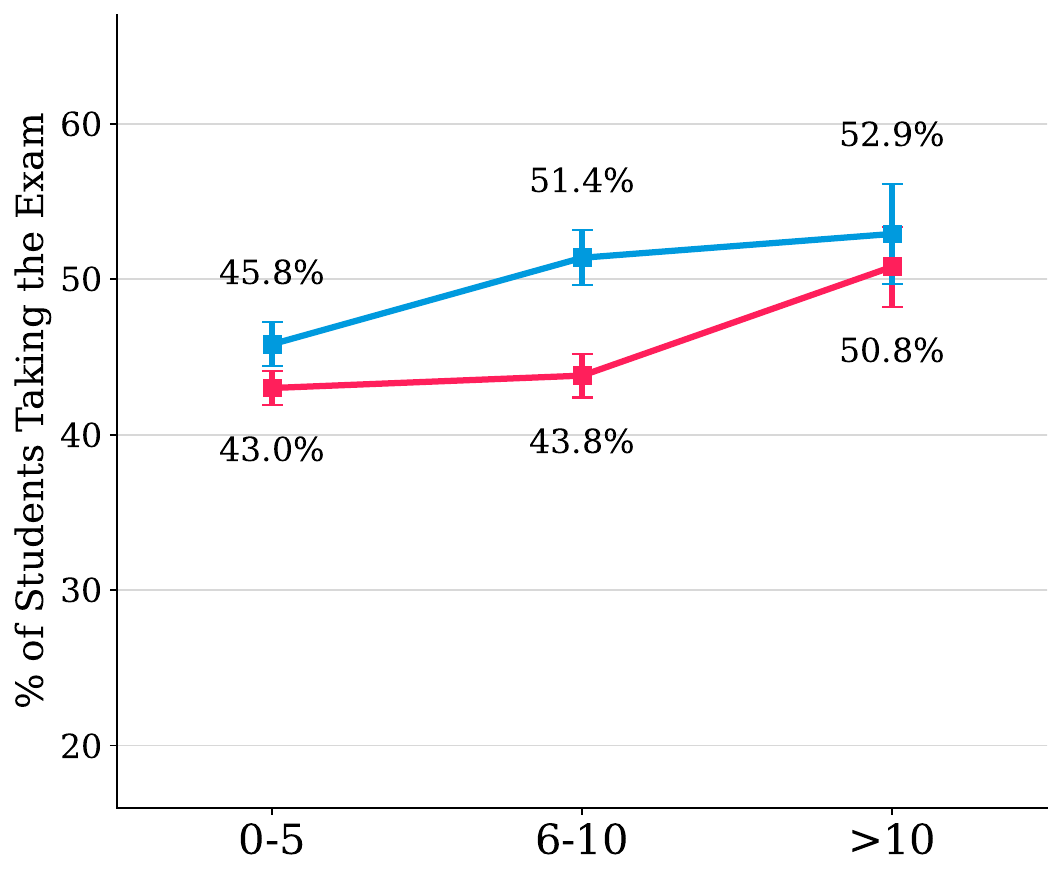}
        \caption{Student Prior Coding Experience}
        \label{fig:exam_experience}
    \end{subfigure}
    \caption{We present exploratory analyses for the experiment-control exam participation gap by looking at different student demographics or country information. The numbers are the average exam participation rate in the two groups. (a) The trend was reversed for the United Nations Human Development Index (HDI). Students from countries with low HDI (<0.55) participated in the exam more, but students from countries with very high HDI  ($\geq$0.8) seemed to participate less. The categorization of HDI is adopted from the UN standard. (b) A large difference is for students who have a medium amount of prior coding experience. (c) A large difference in exam participation exists for students between 23-40. We include analyses for other student covariates in the appendix. The vertical line is the standard error.}
    \label{fig:exam_by_covariates}
\end{figure}

\subsection{Advertisement of GPT-4 Leads to Lower Exam Participation Rate}
\label{sec:low_engagement}

In Code-in-Place, student activity is tracked in three ways: their submission of weekly homework, their attendance for weekly sections, and whether they watched video lectures or interacted with the course materials. The diagnostic exam is voluntary, though students who take the exam receive an additional distinction on their course completion certificate. Among all activities, the diagnostic exam and homework demand the most time.

We found that only 44.1\% of the students in the experiment group took the diagnostic exam, compared to 48.5\% of the students in the control group who took the exam (Figure~\ref{fig:engagement_hist}), representing a significant difference in means ($\Delta$=-4.3pp, SE=1.3, 95\% CI=[-6.9, -1.7]). This effect is very highly significant using a traditional t-test (Unadjusted $P$=.001) and is still highly statistically significant after applying  Bonferonni correction\footnote{
We used Bonferroni corrections to account for multiple hypothesis testing. To apply Bonferroni correction, we multiplied the $P$-value by 15, representing the number of hypothesis tests we conducted in this study.} ($P$=.020).
Similarly, we found a significant difference in means for week 6 homework completion rate between the two groups ($\Delta$=-4.6pp, SE=1.3, 95\% CI=[-7.2, -1.9], $P$=.01, Unadjusted $P$<.001),  though initially there was no discernible difference between students in the experimental and control groups before week 4 (the start of the experiment). Figure~\ref{fig:engagement_curve} shows the students' weekly homework completion rate for the experiment and control group: note a homework is considered ``complete'' when a student has solved all the problems it contains, which range from 1 to 6 per week. 

Although the trend of students engaging less with course activities that require more time commitment is common~\citep{kizilcec2013deconstructing}, offering students novel technology leading to less engagement is highly uncommon because the novelty effect has long been established in education research as a strong factor in increasing student engagement~\citep{makel2014facts}. In fact, in the same course, all other experiments with the same advertisement mechanisms have produced a result where the novel feature drives up engagement~\citep{markel2023gpteach,demszky2023can,bigman2021pearprogram}. Such a \textit{decrease} in engagement is surprising.

This trend of disengagement has some surprising interactions with student demographics. As part of the student application process, we ask them to self-report some demographic information, including their age, gender, country of residence, prior experience with coding, and English fluency. Since we assigned the students randomly to the experiment and control group, examining the engagement difference conditioned on demographic features will not bias the outcome measurement. The general distribution of these covariates can be found in Table~\ref{tab:demographics}.

\paragraph{Engagement Increase For Low HDI Countries:}
The United Nations computes a Human Development Index (HDI) for each country every year. HDI is a measure that summarizes the average health, knowledge (schooling and education), and standard of living for people living in that country. HDI is often used as a metric that describes the level of development for countries (less vs. more developed countries). 
Prior work on MOOCs has found that students in countries with higher HDI tend to earn certificates in MOOCs at a higher rate than students in lower HDI countries~\citep{hansen2015democratizing},  and prior studies have worked to close this gap~\citep{kizilcec2017closing}. Interestingly, we found that offering students access to GPT-4 seemed to increase the student exam participation rate
for students from low HDI countries. There are 325 students from 16 low HDI countries in Asia (Pakistan, Yemen, Afghanistan), North America (Haiti), and Africa (Nigeria, Rwanda, Ethiopia, Tanzania, Sudan, Mozambique, Mali, Madagascar, Uganda, Benin, Gambia, Senegal). 180 students remained active after week 1. In our trial, 111 students were put in the experiment group (61.7\%) and 69 students in the control group. In the experiment group, 16 students (14.4\%) interacted with GPT-4 -- almost identical to the GPT-4 usage for the general student population (14.2\%).

Surprisingly, the disengagement trend reversed for the students who are from the low HDI countries: the exam participation in the experiment group is notably higher at 42.3\% compared to participation in the control group, at 27.5\% (see  Figure~\ref{fig:exam_hdi}). The effect size is small to medium  (Cohen's d=0.31), which is notable. However, our sample size is small, and the raw Student's t-test (Unadjusted $P$=.045) is not statistically significant after adjusting for multiple hypothesis tests ($\Delta$=14.8pp, SE=7.2, 95\% CI=[0.6, 29.0], $P$=.680). We also computed the week 6 homework completion rate for students from low HDI countries and found a similar trend.
Still,  these results are especially encouraging as there is an ongoing effort in the global education community to use GPT-4 to boost educational resources in underserved areas due to unrest, war, poverty, or extreme events.~\citep{butgereit2023prof,choi2023llms}.

\paragraph{Age and Coding Experience Suggest a ``Middle Experience Gap'':}

Student disengagement may also mediated by age (Mean=31.4, SD=10.4, Median (IQR)=29.0 (23.0-37.0), Max=84.0) and prior coding experience (Mean=5.1, SD=4.4, Median (IQR)=4.0 (2.0-8.0), Max=18.0). We divide the students based on their age into four groups: 18-22 (college age) (N=1,241) and then 20s (N=1,947), 30s (N=1,588), and 40s and above (N=1,055). A difference between exam participation is suggested but only for students between 23-40 (Figure~\ref{fig:exam_age}).  We discuss why this might occur in Sec~\ref{sec:discussion}. A critical factor to note is that students take Code-in-Place for different reasons, and job eligibility is a particularly common reason for students aged 23 to 40. 
We notice a similar phenomenon for students with different prior coding experiences, which we ask students to self-report and rate them on a scale of 0-18. We divide students into three groups: no experience 0-5 (N=3,180), less experienced 6-10 (N=2,037), and more experienced >10 (N=614). We notice a significant decrease in engagement for students in the less experienced group ($\Delta$=-7.5pp, SE=2.3, 95\% CI=[-12.0, -3.1], $P$=.014, Unadjusted $P$<.001). We show the difference between the experiment and control group among students with different coding experiences in Figure~\ref{fig:exam_experience}.

\paragraph{Other Demographics}
We did an exploratory analysis of other demographics, such as English fluency and gender. English fluency is correlated with HDI (\textit{r}=0.10), and similar to prior coding experience, we ask the student to self-report their fluency in English on a scale of 0-20  (Mean=13.8, SD=2.5, Median (IQR)=14.0 (12.0-16.0), Max=20.0). We then divide students into two groups: not fluent $\leq$ 10 (N=551), and fluent > 10 (N=5,280). We see no exam participation difference between the experiment and control group for students who are less fluent in English and a large difference for students who are fluent (Figure~\ref{fig:exam_fluency}). We did not see an interaction of gender with exam participation rate change beyond the general trend of the experiment group having a lower participation rate than the control group (Figure~\ref{fig:exam_gender}).

\subsection{Estimated Positive Benefit of GPT-4 on Learning For Adopters}
\label{sec:exam_score}

\begin{figure}[t]
    \centering
    \begin{minipage}{0.43\textwidth}
        \centering
        \includegraphics[width=\linewidth]{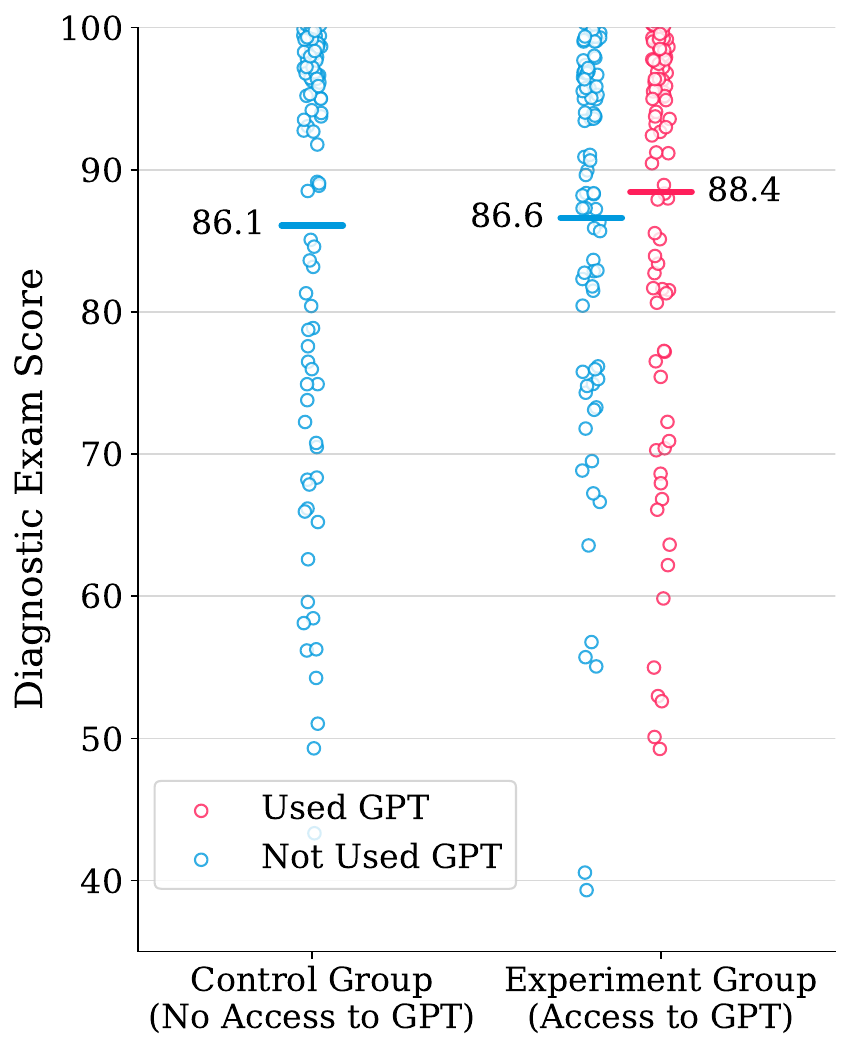} %
        \caption{We plot the diagnostic exam score distribution for students in the control and the experiment group. For the experiment group, students who used GPT-4 before the exam are adopters. We plot students who did not use GPT-4 as blue dots and students who used GPT-4 as red dots. The bar shows the average over the entire population.}
        \label{fig:control_exp_score}
    \end{minipage}\hfill
    \begin{minipage}{0.55\textwidth}
        \centering
\small

\begin{tabular}{@{}lccc@{}}
\toprule
&  \begin{tabular}{@{}c@{}}$\Delta$ \textbf{Exam} \\ \textbf{Score}  (pp)\end{tabular} & 90\% \textbf{CI} & \begin{tabular}[c]{@{}c@{}} \textbf{ES} \end{tabular} \\ \midrule
\multicolumn{4}{l}{\textbf{E1}: Advertisement Effect}   \\ \midrule
Ignore Missingness                           & \multicolumn{1}{c}{+0.91} & [-0.00, 1.89]  & 0.05        \\ %
Impute for Missingness                                                                & \multicolumn{1}{c}{+0.67} & [-0.37, 1.72]   & 0.09      \\ \midrule
\multicolumn{4}{l}{\textbf{E2}: Effect for Adopters}   \\ \midrule
Ignore Missingness                                                                        & \multicolumn{1}{c}{+4.49} & [-0.31, 9.66] & 0.23          \\ %
Impute for Missingness  & \multicolumn{1}{c}{+6.86} & [0.30, 14.13] & 0.40         \\ %
\bottomrule
\end{tabular}

\captionof{table}{\textbf{Estimated Exam Score Improvement}: We report the estimated difference on exam scores, a 90\% CI on that difference, and an effect size (ES). For \textbf{E1}, we report the score difference between students who were given access to GPT and students who were not. %
For \textbf{E2}, we report the estimated average treatment effect on exam scores (due to use of the LLM)  for adopters, students who would choose to use GPT-4 if provided access in the course. 
We estimate adopters have a \textbf{6.8  percentage point} average treatment effect improvement in their exam scores due to using the provided LLM.} %
\label{table:learning_gain}
    \end{minipage}
\end{figure}

In the previous section, we observed that providing access to GPT-4 had a negative impact on overall student \emph{engagement} in this course. In this section, we estimate the impact of GPT-4 access and usage on student learning \emph{outcomes}, as measured by exam scores.
Specifically, we compute two separate estimates:
\begin{enumerate}[itemsep=0pt, wide=0pt, leftmargin=*, after=\strut,label={(\textbf{E\arabic*}):}]
    \item \ittEffect: Is there a change in exam performance due to advertising and offering students access to GPT-4? This effect can be measured by computing the difference between the means of student exam scores in the experiment and the control group, accounting for missing data. 
    \item \treatEffect: Is there a change in exam performance for \textbf{adopters}, the students who chose to use GPT-4 in our experiment group? Intuitively, these adopters might be more active and more engaged with the class. Allowing them to use a new tool might lead to a bigger impact on their learning activities. We use a causal estimator to compute this effect.
\end{enumerate}

It would also be ideal to understand the impact on learning gain for the broader student population, not just the adopters (who may be a different distribution of students). However, given that we didn't require students to use GPT-4, we cannot estimate this effect given our experiment. 
In the following sections, we first describe the general distribution of the student exam scores. Then we discuss how we can use causal inference to estimate the two effects of interests ($E_1$ and $E_2$) and how to handle missing outcomes from students who did not take the diagnostic.

\paragraph{Distribution of Student Exam Scores}
We plot the diagnostic exam scores of the students in the control and experiment groups separately and use two colors to indicate which student used GPT in Figure~\ref{fig:control_exp_score}. Many students achieved a perfect score on the exam. The students who took the exam in the experiment group (48.5\%) had an average exam score of 87.0 (N=1,579, SD=18.7, Median (IQR)= 95.9 (80.8-98.6), Max=100.0). The students who took the exam in the control group (44.1\%) had an average exam score of 86.1 (N=1,089, SD=20.0, Median (IQR)= 95.9 (79.5-98.6), Max=100.0). We could give students who didn't take the exam a score of 0. This is problematic because it is highly likely that there are confounding reasons that impact both why a student takes the exam and their performance: for example, students who have time to take the exam also might have more time for the course generally. A zero score is almost definitely an underestimate of the potential score of any student who did not take the exam. Note that using zero as a substitute for missing data will also significantly lower the average score for the group that had more missing outcomes. Imputing missing exam scores as zero, the students in the experiment group have an average exam score of 38.4 (SD=44.9, Median (IQR)= 0.0 (0.0-93.2), Max=100.0). The students in the control group have an average exam score of 41.7 (SD=45.2, Median (IQR)= 0.0 (0.0-94.5), Max=100.0).

\paragraph{Handling Missingness}
There are two common strategies for addressing missing outcomes in a randomized control trial that rely on different assumptions. When data is missing completely at random (MCAR), which in our setting would imply that it is completely random whether students choose to take the exam or not (regardless of experiment condition),  then missingness can be ignored, and the causal effect can be estimated using only the population who took the exam. We report this strategy as ``Ignore Missingness,'' and it is valid under the MCAR assumption. MCAR is a strong assumption and unlikely to be valid in our setting, as we expect there to be differences among students who do or do not take the exam. A second common assumption is ``missing at random,'' which is that the data is missing at random conditioned on the observed covariates. In our setting, this implies that given our observed features for each individual, such as participation in the class early in the course, or static demographic variables, such individuals with identical covariates will take the exam with equal probability. When data is missing at random, it is common to use imputation to estimate missing outcomes. In our setting, this involves using a machine learning model to estimate the exam scores of the students who didn't take it. This method may introduce bias to the estimated causal effect from the learned imputation model. We use cross-fitting, a strategy to avoid overfitting and increase data efficiency, to estimate our causal effect. Cross-fitting involves using one part of the data to estimate the nuisance parameters of the imputation model and the other part of the data to estimate the causal effect. Such ideas have been introduced to build \emph{honest} causal trees~\citep{athey2016recursive} and train double machine learning models~\citep{chernozhukov2018double}. We perform automatic model selection and cross-validation to reduce subjective bias. We report our procedure in the Appendix. We refer to this method as ``Impute for Missingness'' in Table~\ref{table:learning_gain}.

\subsubsection*{Exam Score Increased For Adopters}

14.2\% of students who were given access to GPT-4  used GPT-4. We refer to these students as the adopters. \cite{rogers2014diffusion} estimated that the percentage of people in a population who are willing to experiment with new technology is roughly around 13.5\%-16\%. We show the distribution of the scores of students who used GPT (in red) and who did not use GPT (in blue) in Figure~\ref{fig:control_exp_score}. We are interested in knowing if there is a change in exam scores due to GPT-4 usage for adopters. We analyze this in three ways. We first look at the exam performance of students in the control and students in the experiment group who did not use GPT-4.  We conducted a hypothesis test on the difference of means between the two groups and found the difference between the two groups is very small and not statistically significant when we ignore missingness ($\Delta$=0.54pp, SE=0.81, 90\% CI=[-0.79, 1.87], Cohen's $d$=0.03, $P$>.99, Unadjusted $P$=.504). Our finding is similar when we impute for missingness using our regression model ($\Delta$=0.43pp, SE=0.38, 90\% CI=[-0.19, 1.05], Cohen's $d$=0.03, $P$>.99, Unadjusted $P$=.247). This suggests that the advertisement has little impact on exam scores if students do not subsequently use GPT-4. 

We next estimate if there is a difference in exam performance between the adopters (students in the experiment group who used GPT-4 before the diagnostic exam) and the students in the control group. 
The difference has an effect size that falls within the small range, but we are not certain if it is statistically significant after adjusting for multiple hypothesis testing ($\Delta$=2.36pp, SE=1.14, 90\% CI=[0.49, 4.24], Cohen's $d$=0.13, $P$=.836, Unadjusted $P$=.056). When we refill the missingness with a regression model, we found the difference to be larger ($\Delta$=2.89pp, SE=0.69, 90\% CI=[1.76, 4.02], Cohen's $d$=0.21, $P$<.001, Unadjusted $P$<.001).
The difference between the adopters and all the students in the control group does not equal \textbf{E2} (the improvement in exam performance for adopters). 
To see this, imagine that 14.2\% of students in the control group are also adopters—they couldn't use GPT-4 because we didn't offer them access. The exam score improvement should be computed by the difference between the adopters in the control group and the adopters in the experiment group. Unfortunately, we don't know who the adopters are in the control group.

While we would like to estimate the average treatment effect on exam performance, it is likely that adopters who used GPT-4, when offered, are significantly different than other students. 
\cite{imbens1994identification}'s seminal work proved that the average treatment effect on compliers, in our case, the adopters who started using GPT-4 when offered, can be estimated assuming that (1) the advertisement impacts student usage of GPT-4 (2) the advertisement's impact on student exam scores is solely through whether the student uses GPT-4, and (3) there are no students who would have used GPT-4, but did not, because they received the advertisement. We can verify assumptions one and three hold. The second assumption, also known as the exclusion principle, is impossible to verify and relies on domain experts to judge if it is reasonable. In our setting, we believe it is reasonable, and we consider this further in the discussion. Imbens and Angrist named the treatment effect on compliers (in our case, adopters) as the local average treatment effect (LATE).

We computed the local average treatment effect (LATE) for \textbf{E2}. The standard Cohen's $d$ formula cannot be used to compute the effect size for a LATE estimator. We use the proposal from \cite{bansak2020generalized} to estimate the effect size (ES). We compute the impact on exam performance for adopters and impute for missingness using the machine learning model.  The estimated LATE is positive, at $\Delta$=6.86.  Though the difference is not statistically significant after adjusting for multiple hypothesis testing (BCa 90\%CI=[0.30, 14.13], ES=0.40), the estimated effect size is near medium\footnote{The use of effect sizes in assessing educational interventions is nuanced: \cite{kraft2020interpreting} discusses a number of features that impact effect sizes, and he makes specific suggestions for lowering the category thresholds in education for K-12 when measuring impact on learning.}, at  $0.40$, suggesting the potential of a substantial positive effect on exam scores on adopters.\footnote{For completeness, we also computed a LATE estimate where we ignore the missingness (by excluding students who did not take the diagnostic exam). This effect had the same trend, though it was slightly smaller: ($\Delta$=4.49, BCa 90\% CI=[-0.34, 8.98], ES=0.23).}  %

\subsection{GPT-4 Usage Behaviors}
\label{sec:usage_analysis}

\begin{figure}[ht]
    \centering
    \includegraphics[width=0.6\textwidth]{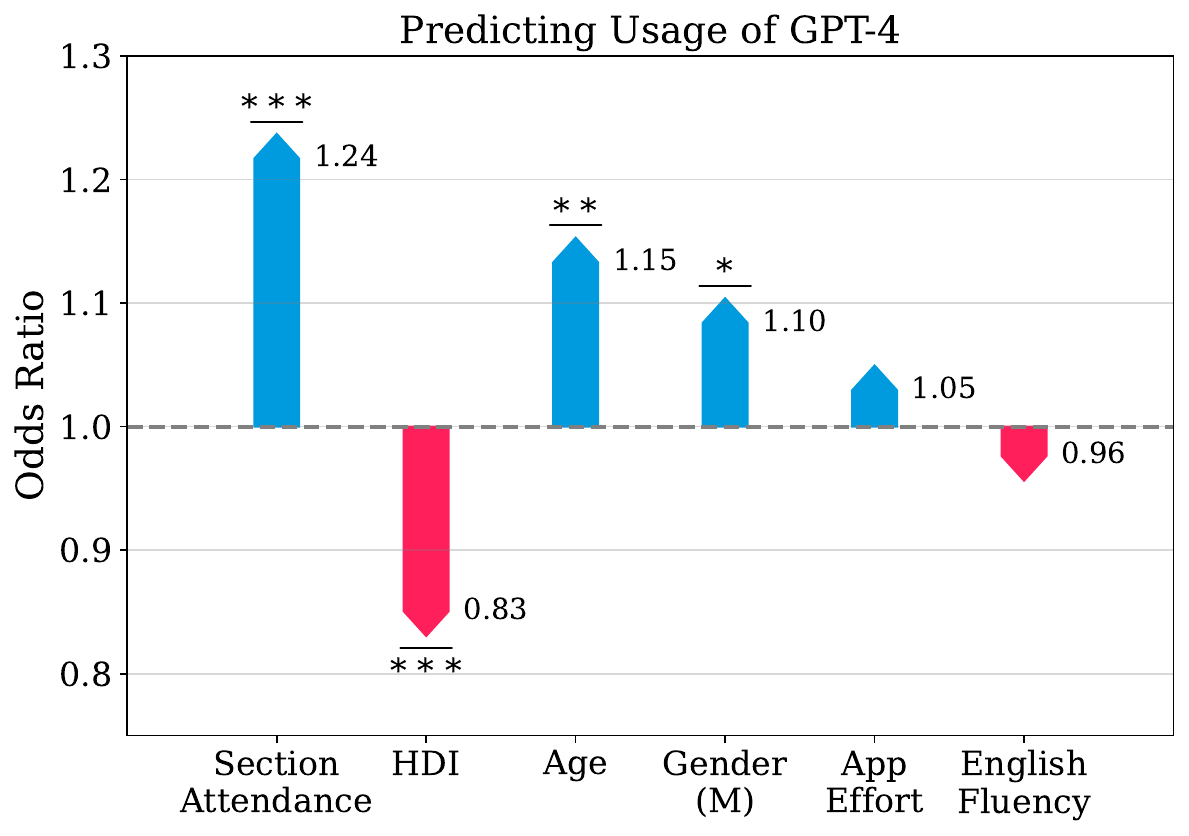}
    \caption{\textbf{Student Demographics and GPT Usage}: We use a logistic regression analysis to understand what type of student is more likely to use GPT if offered. We report the odds ratio (i.e., exponentiated logistic regression coefficients) for each demographic feature. We include the full report in the Appendix. ***$P$<0.001, **$P$<0.01, *$P$<0.05.}
    \label{fig:gpt_usage}
\end{figure}

In order to understand the impact of offering and using GPT-4, we analyze student usage behaviors in two ways. First, we analyze the characteristics of the students who are using GPT-4. 
We used logistic regression to predict if a student would use GPT-4, given a set of prior student covariates, using data from the experiment group. We report the logistic regression coefficients in Table~\ref{tab:usage_regression}, and we visualize a subset of their odds ratio (the exponential of the coefficients) in Figure~\ref{fig:gpt_usage}.

We can see that naturally, students who attended more sections prior to week 4 (i.e., these students have been very active in class) are more likely to use GPT-4 if offered to them. If they were observed having attended all three sections in prior weeks, their odds of using GPT-4 would have increased by 24\%. For HDI (the HDI is an index normalized to be between 0 and 1), students from the most wealthy country (HDI=1) will have 17\% lower odds of using GPT-4 than students from the poorest country (HDI=0). Similarly, we observe a correlation between GPT-4 usage and age and gender as well. Slightly older students are more likely to use GPT-4. Male students are more likely to use GPT-4 than female, non-binary, and students of other genders. We did not find statistical significance in other characteristics, including English fluency.

\begin{figure}[ht]
    \centering
    \includegraphics[width=0.9\textwidth]{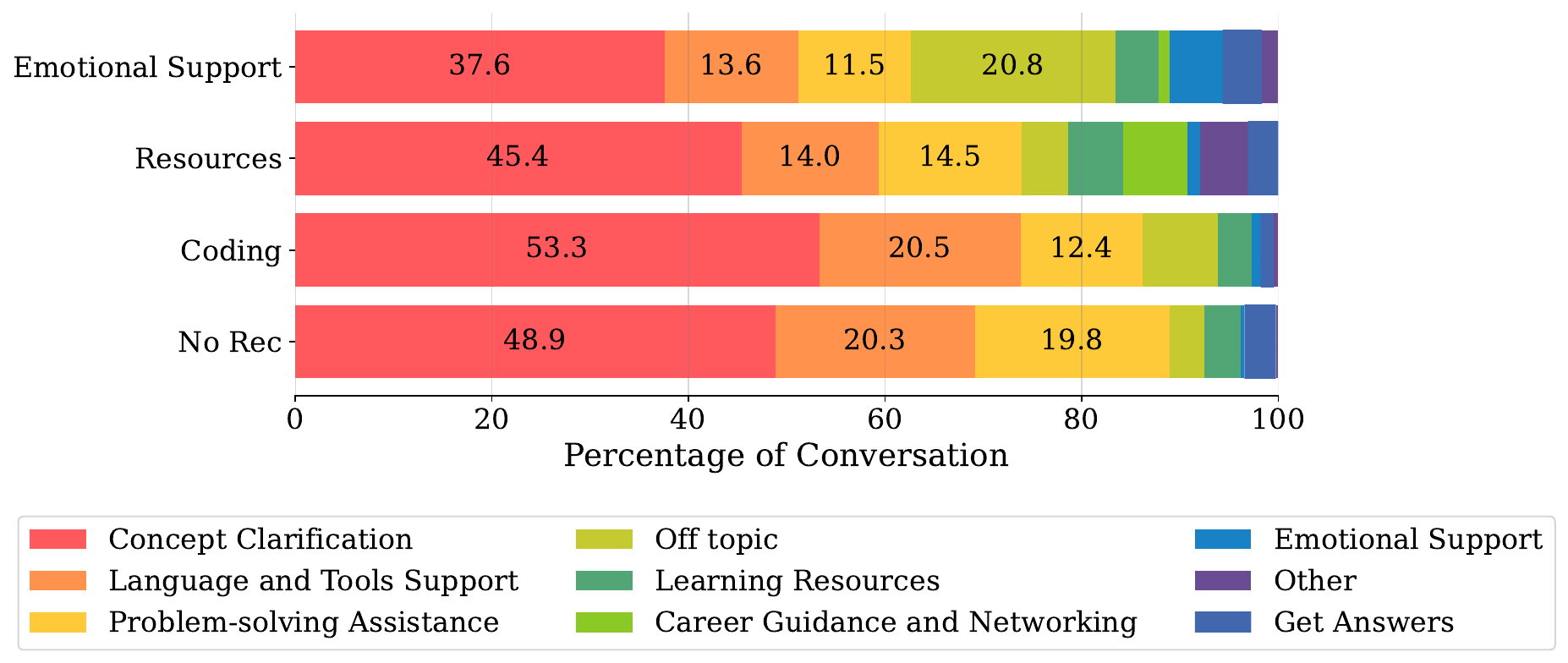}
    \caption{\textbf{Student-GPT Conversation Topics}: We randomly assigned students to 4 nudge groups where we showed them some potential conversation starters with LLMs. Under each group, students engaged in slightly different conversations. We display the percentage numbers for each block if the percentage > 10\%. ``Get Answers'' is a category where students tried to trick GPT-4 into giving them correct answers to the homework (1-4\% across conditions).}
    \label{fig:gpt_usage_pattern}
\end{figure}

We also investigated what kind of conversation the student was having with GPT -4 in our class. The students in the experiment group were randomly (with equal probability) put into four groups, where they were shown some suggested questions that they could ask GPT-4. We created three sets of potential questions, broadly categorizing them as ``Coding'', ``Resources'', and ``Emotional Support''. We manually wrote 10-20 questions in each of these sets. Students who got assigned to these groups will see three randomly sampled suggested questions every time they open the custom-built interface (See Figure~\ref{fig:interface} (b)). We also created a group called ``No Rec,'' where the students will not see any suggested questions. We created this group to study what the students would naturally ask in a coding class setting. 

We see that most students who reach out to GPT-4 do so to ask questions related to coding concept clarification or the specifics of the programming language they are learning. The percentage of concept clarification ranges from 37.6\% to 53.3\%, and language and tool support ranges from 13.6\% to 20.5\%. A majority of these discussions stayed on topic. We only found 1-4\% of conversations are directly related to students trying to deceive the GPT-4 to give them answers to the homework question. We have pre-emptively added homework assignments as prompts so that GPT-4 would not easily give students direct answers. However, we did not spend enough time testing whether it was easy to bypass our safeguards. After examining the transcripts, we did not find students asking diagnostic exam questions through the GPT-4 interface. However, we are not able to verify whether students have used OpenAI's public ChatGPT interface outside the class.

One encouraging finding is that, by displaying different types of example questions, we are able to induce different behavior patterns from the students. The percentage difference of the same topic between different groups can be as high as 15.7\% (e.g., ``Concept Clarification''). Our experiment period was only 9 days. Therefore, we cannot conclude whether this difference will disappear once students have more time to use GPT-4. However, this does suggest that small nudges may impact how students interact with the LLM for educational purposes. We did not find any educational outcome difference between these groups.

\section{Discussion}
\subsection{Limitations}
\label{sec:limitations}

\textbf{A Particular Course, with Particular Students, at a Particular Moment in Time}

A common limitation of educational research, even randomized control trials, is that experiments are conducted in a particular context. In this case, we ran the student in a particular domain (computer science) with a particular cohort of students at a particular moment in time. Consequently, the findings from this study only directly apply to this particular context.  The same intervention, if studied in different learning environments, may produce different results. Our course's unique characteristics likely shape the educational experience and outcomes observed: 
(1) The course is at-will in contrast to many k-12 or university experiences where learners must complete the whole experience in order to get a grade and/or credit. Students in traditional classrooms may have experiences that could lower their motivation, but they are less likely to drop-out. 
(2) The course centers on human teachers. As such, it likely attracts learners who are interested in education from humans. This may make our learner population different than that of a typical massive online course.
(3) The impacts might be specific to programming education. Some of the causal mechanisms for why students disengage may play out differently in courses on literature, history, or even math. This is especially believable because coding is a computer-oriented domain in which LLMs particularly excel.
(4) The experiment was conducted in 2023, a period marked by a distinctive zeitgeist regarding artificial intelligence in education. While perceptions and news differ by geographic location, broadly, the news was a mixture of excitement about the potential of large language models and fears over the role that LLMs may play in society.

These limitations are typical in the field of education. We urge the reader to be cautious about extrapolating these results into different contexts.  Despite these contextual limitations, a large-scale randomized control trial case study remains one of the most useful ways of knowing within the educational domain. The insights gained, while specific, can be a helpful data point for those trying to pursue a broader understanding of educational interventions and their impacts. 

\textbf{User Experience Design May Impact Results}

Several decisions regarding user experience were implemented that may have influenced the study's outcomes. Firstly, the experiment was advertised through email and a link from a main tab on the homepage rather than being integrated within the IDE or the learning experience itself. Second, the interface was purposely designed to resemble the ChatGPT user interface but was differentiated by varying color schemes to clarify to students that they were navigating a distinct environment. Finally, before students used the chat experience, we chose to show them a brief handout detailing the advantages and disadvantages of Large Language Models (LLMs). We made the decision to include the short handout in the interest of making sure our students were informed (see figure \ref{fig:interface}). We included the positive and negative versions to better understand how a teacher's stance could impact usage. We observed no significant difference in long-term engagement levels between the group that received positive emphasis and the group that emphasized cautionary points. The lack of a difference might indicate that the content of this handout did not have an impact on the outcomes observed. On the other hand, perhaps the existence of any handout, even if positive and short, could have been a point of friction that impacted outcomes. 

These design choices are worth considering, especially due to the 14\% adoption rate among students. Though 14\% is low for a mandatory feature in a course, it is a rather typical adoption rate for an optional tool in this particular course. For comparison, a second optional tool in the course, designed to provide style feedback and also advertised via email, experienced an even lower participation rate of 7\%, as detailed in \cite{woodrow2024ai}. While stronger encouragement (or even making use of the tool mandatory) could drive up engagement in the tool, it could exacerbate the decrease in engagement in the course.

\subsection{What Caused Disengagement?}
\label{sec:discussion}

In the randomized control trial, there was a notable difference in engagement on a two-week time scale between learners with and without access to ChatGPT. This effect is influenced by student age, coding experience, and HDI of the student's country. We note that it is surprising to see a drop in engagement in education simply from getting access to an optional tool. Many learning science experiments are ``doomed to succeed'' because of the strong novelty effect in education \citep{makel2014facts}. This novelty effect is seemingly non-existent when providing an LLM chat interface to students.
On their own, these results do not suggest  \emph{why} there was a difference in engagement. In this section, we present a few theories on such a causal mechanism. 

\textbf{Job Threat Hypothesis}

One hypothesis for the impact on engagement is that access and advertisement to GPT confronted learners with a threat to their prospects of getting a programming job. Several studies have shown that learners who make a connection between
academic courses and activities and future versions of themselves
experience increased motivation towards their learning \citep{leondari2007future, lens2001student, greene2004gender}, especially in intro to programming courses. Specifically, \cite{peteranetz2016perceived} showed that a difference in a student's perception of an instrumental connection between course participation and employment was a significant predictor of standardized course grades. Under the Job Threat Hypothesis, interacting with ChatGPT decreases a student's instrumental connection between learning and obtaining employment. When the student either uses ChatGPT or reads the advertisement, the student supposes that there may be fewer jobs and there will be steeper competition for those limited roles. At the very least, a student might perceive a greater uncertainty around jobs.

This hypothesis is supported by the observation that the biggest engagement impact of access and advertising of GPT was on people who demographically match those who are applying for jobs -- people in the job-seeking age (22-40) and middle-experience programmers. The instrumental connection between learning to code and jobs is especially pertinent in the Code in Place course. Of the students in the class, 46.8\% of students list, ``I want to get a job as a programmer'' as a motivation for taking the class upon submitting their application. The percentage of students who listed jobs as motivation is even higher for learners of job-seeking age (49.0\%). In contrast, only 43.7\%  of learners outside that range listed jobs as a motivation. As a corollary, there was a 1.26 percentage point decrease in exam participation among learners in the experimental condition who listed a job as a motivation for taking the course (43.4\% exam participation) compared to those who did not list jobs (44.7\% exam participation).

\textbf{AI Mistrust Hypothesis}

Artificial intelligence's integration into society has engendered polarized opinions. IPSOS ran a global study of perceptions of AI at the same time that the Code in Place experiment ran, where they interviewed 22,816 adults under the age of 75 across 31 countries \citep{IpsosGlobalAI2023}. They found that 52\% of respondents were nervous regarding AI, and 46\% did not agree that there were more benefits than drawbacks. These broad phenomena appeared to be playing out inside the course. In conversations with learners in Code in Place, we noticed a surprisingly high level of distrust in AI. Students voiced concerns which range from concerns over (1) privacy, (2) concerns about the trustworthiness of AI, as well as (3) more existential worries about the role of AI in society \citep{kochhar2023us}. As such, as evidenced by the IPSOS survey and conversations with our students, it is reasonable to assume that a subset of our learners (large but of unknown size) do not trust AI. We also assume that there is a separate subset of learners who are excited about the potential. We hypothesize that those with a negative association towards AI are more impacted by a teaching chatbot than those with a positive attitude. As such, the polarization surrounding AI, fueled by these concerns, may contribute to a decrease in engagement among students when AI tools are introduced into the learning environment. The country-level variance in trust seems to support this hypothesis. In the IPSOS survey, they found that ``Trust in AI varies widely by region; it is generally much higher in emerging markets ... than in high-income countries.'' This ordering of countries by IPSOS trust in AI reflects the ordering by country of effect-sizes that the advertisement of GPT had on students. Specifically, we also note that learners from emerging markets (largely those with lower HDI) were positively impacted by access to ChatGPT.

People are more comfortable with AI automating mechanical tasks than human or relationship labor, such as caring for children. Where does teaching fit into this spectrum? There is a broad set of literature that argues, ``The need for social belonging—for seeing oneself as socially connected—is a basic human motivation'' and as such, ``
social connectedness ... has a dramatic impact on course completion'' \citep{walton2007question}. Providing more AI assistance likely induces fewer human-to-human interactions. Could this reduction in interpersonal engagement lead to feelings of loneliness and alienation? These concerns are particularly potent in courses like Code in Place, designed to emphasize a human-centric learning experience. In such environments, the introduction of AI tools such as ChatGPT might be perceived as especially intrusive or displacing, contributing to a decrease in student engagement.

\textbf{AI vs Human Identity Threat}

The dosage of help from an AI may be inducing a novel form of identity threat. 
While learning, students interpret their successes and difficulties as rewards that influence their identity and self-perception as burgeoning coders \citep{kaplan2012identity}. This self-reflection can be especially modulated by demographic identity in a mechanism often called ``identity threat'' \citep{hanselman2014threat}. While traditional identity threat focuses on gender and ethnicity, we may be observing an identity threat oriented around simply being human.  When introduced to a powerful, capable assistant like ChatGPT, students might be confronting the stressful idea: ``I find coding challenging, yet here is an AI that excels effortlessly.'' Even if jobs are not a core goal for the learner, this could impact their self-perception as a ``coder.'' Identity has a causal impact on motivation, which could explain the drop in engagement \citep{cohen2008identity}.

\textbf{Learners Found a Better Way to Gain Skills}

An alternative and somewhat more optimistic hypothesis for the observed decrease in engagement posits that advertising ChatGPT introduces students to a better opportunity for reaching their learning goals than the course: directly engaging with ChatGPT. Imagine a student who was unaware of ChatGPT and the recent advances of Generative AI. After the advertising of the Code in Place deployment of ChatGPT, they interact with the LLM and have a productive conversation. In order to circumvent the protections placed to stop learners from obtaining solutions to course challenges, they switch to using the OpenAI website directly. There, they are able to learn at their own pace in an environment that is adaptive to their particular needs. They no longer have a strong need to return to Code in Place as their education needs are being met. Even though these students are achieving their learning goals, we have lost our ability to observe their improvement, and they are categorized as having a drop in engagement.

This shift in perception could be especially pronounced for students who are self-motivated or who prefer autonomous learning environments. They might see ChatGPT not just as a supplement to their learning but as a complete alternative to traditional educational pathways. This hypothesis is supported by observing that students had productive conversations with ChatGPT and seemed to be getting pedagogical value out of their chats.

\subsection{What Caused Improved Exam Scores?}

Our analysis of the adopters' interactions with GPT-4 shows that those who used GPT-4 were generally using it in a constructive way to better their understanding of the material and the field at large.  This could have improved their understanding, which transferred to test performance gains. An interesting question for future study is how the type of use of GPT-4, and the amount, might impact learning outcomes. This particular result is less surprising. There is a long-held belief in education that direct one-on-one tutoring substantially improves student ability \citep{bloom19842}. LLM chat seems to be a reasonably useful autonomous tutor. One concern was that access to an AI tutor at any point could serve as too much help, sometimes labeled the ``Scaffolding Paradox'' \citep{gillespie2017paradox}. Instead of grappling with problems, iterating through solutions, and learning from errors-- a crucial cycle in developing programming acumen-- students may shortcut this process by seeking immediate answers from AI, thereby steepening their learning curve in the long run.  Fortunately, the strong exam performance of students who adopted ChatGPT suggests that overreliance on AI, to the detriment of learning, was not a dominant mechanism in our setting.
\section{Related Work}

Recent studies have explored the effects of large language models on education. One branch of work focuses on making observational and qualitative studies of how LLMs are used in different educational settings. \cite{butgereit2023prof} launched a WhatsApp bot that connects to GPT-4 to tutor mathematics at the university level in the Arabic language, particularly for areas going through periods of violent unrest. \cite{choi2023llms} collected data on how teachers from the poorest schools in Sierra Leone used GPT-4 to assist their teaching. They found that 48\% of the questions asked to GPT-4 were about concept clarifications, similar to our finding as well. \cite{liu2024teaching} integrated GPT-4 into a university-level class (CS50) and found that students enjoyed the experience and found LLMs to be helpful.

Another branch of work focuses on designing randomized control trial (RCT) experiments to understand the effect of LLMs on educational outcomes. \cite{kumar2023math} conducted a large-scale experiment and found that GPT-generated hints positively impacted learning compared to not providing hints at all. The participants in this study were crowd-sourced site workers rather than individuals from a traditional classroom or realistic educational environment. \cite{prihar2023comparing} showed that GPT-generated hints have lower qualities compared to human tutor-written hints when rated by other teachers. \cite{pardos2023learning} showed that when students are using these hints for learning, learning gains were only statistically significant for human tutor-created hints.
As far as we know, there hasn't been any RCT work directly examining the potential learning benefits of introducing students directly to ChatGPT in a classroom setting.

\section{Method: Local Average Treatment Effect}
\label{sec:method}

There is a long history in economics and social sciences of estimating causal effects by using instrumental variables to account for possible self-selection into a treatment. Using this strategy, economists have been able to estimate that there is a  negative impact on lifetime earnings when people are drafted, and gain evidence that increasing years of education can result in higher salaries~\citep{angrist1990lifetime,card1993using,angrist1996identification}. Even though we want to report the expected (average) treatment effect for any person who receives the treatment, due to the self-selection effect, we can only make valid statistical statements about the people whose resulting decision (to uptake a treatment) are impacted by an instrumental variable. To distinguish from the average treatment effect, we report the local average treatment effect -- local in the sense that the treatment effect is averaged over (what the literature refers to as) \textit{compliers}, those who complied with the treatment or control indicated by the instrumental variable (in our case, using GPT-4 after being given access).

\paragraph{Notation} To describe the causal structure of our experiment setup, we introduce the following notations. Each student $i$ is represented by $(X_i, Z_i, W_i, Y_i)$, where $X_i \in \Reals^d$, a vector of observed $d$ covariates (demographic characteristics, pre-enrollment qualification test, general activities in class before the experiment, etc.); $Z_i \in \{0, 1\}$ is the instrumental variable and indicates whether the student is informed of the existence of free GPT access in class via an email and a quick-access button is provided in the homepage sidebar; $W_i \in \{0, 1\}$ describes whether the student uses the in-class GPT-4 interface that we provided; $Y_i \in \Reals$ denotes the observed outcome, the $i$-th student's exam score. Folllowing the potential outcomes framework, we use $Y_i(0)$ and $Y_i(1)$ to the denote potential outcomes under treatment $W$ for individual $i$, so that $Y_i = Y_i(W_i)$. Note the fundamental problem of causal inference is that we only observe one potential outcome for each individual $i$.

We denote each student's decision to use GPT-4 when given access (C stands for compliance) as $C_i = \{W_i(0), W_i(1)\}$, where $W_i = W_i(Z_i)$\footnote{$W_i(0)$ is short for $W_i(Z_i=1)$, which means student $i$ has been given access and nudged to use GPT-4. $W_i(Z_i=1)=1$ indicates the student has been emailed and given access and chose to use GPT-4 -- this student is a compiler in the causal inference terminology. In earlier parts of the paper, to use more familiar terminology beyond the causal estimation community, we used the word ``adopters''.}. We define the exam score improvement (the treatment effect) of GPT-4 for the students who complied as $\gamma$, following the setup from \citet{angrist1996identification}:
\begin{align}
    \gamma &\coloneqq \E\brr{Y_i(1) - Y_i(0) \mid C_i = \text{complier}}.
    \label{eqn:late}
\end{align}
 $\gamma$ is the Local Average Treatment Effect (LATE). 
 The following assumptions suffice for LATE to be identifiable:

\begin{ass}[Uniformly Random Nudges]
   $Z \ind X$ \thlabel{ass:urn}
\end{ass}

\begin{ass}[Relevance]
    $Z \not\!\!\ind W | X$ \thlabel{ass:relev}
\end{ass}

\begin{ass}[Exclusion]
    $Z \ind Y | W, X$ \thlabel{ass:exclusion}
\end{ass}

\begin{ass}[Monotonicity]
For all $x$, either $P(W=1|Z=1,x) \geq P(W=1|Z=0,x)$ or $P(W=1|Z=0,x) \geq  P(W=1|Z=1,x)$. 
\thlabel{ass:monotonicity}
\end{ass}

\thref{ass:relev} to \thref{ass:monotonicity} are standard assumptions for identifying the local average treatment effect~\citep{angrist1996identification,athey2017econometrics}. 
\thref{ass:relev} assumes the instrument has an influence on the treatment variable $W$, and can be verified empirically. \thref{ass:exclusion}, which assumes that the instrument only impacts the outcome variable through observable variables, is not verifiable. Note that our exclusion assumption accounts for the presence of missing variables that may be influenced by the instrument variable and, therefore, is slightly different than the standard LATE exclusion assumption, which is simply  $Z \ind Y | W$.  Typically, an important part of using instrumental variables is to argue, given domain knowledge, whether the exclusion principle holds. In our setting, we believe this is a reasonable assumption, as we expect the nudge/instrumental variable to have little impact on exam performance, given knowledge of whether the student used GPT-4 or not. 

The monotonicity assumption (\thref{ass:monotonicity}, often called the ``no defiers'' assumptions) is that the instrument variable impacts all individuals in a monotone way: for example, if there exist some compiler individuals, whose participation in the treatment group is increased by an instrument, then there cannot exist any "defier" individuals whose participation in the treatment group is decreased by the instrument.  This is not verifiable, but it is a very common assumption and is plausible in our setting.

\paragraph{Example Calculation}

Here is a simple example of how the estimator works. We use $c$ (compliers) to mark students who would become adopters if we offered them access. We use $n$ to mark students who would not use GPT-4 even if we offered them. We have both types of students in the experiment and in the control group. We use $N_{c1}$ to mark the number of adopters in the experiment group and $N_{n1}$ to mark the number of students who did not use GPT-4 when given access. Analogously, we can define $N_{c0}$ and $N_{n0}$ for the control group as well. Note that we know $N_{c1}$ and $N_{n1}$ in our experiment because we have a detailed record of who used GPT-4 or not when they were given access, but we don't know $N_{c0}$ and $N_{n0}$ because we don't know who are the adopters in the control group. Let the total number of students in the experiment group 
be $N_1$ and the total number of students in control $N_0$. 
We use $i \in c0$ to mark students in the experiment group who are adopters (observed) and $i \in c1$ to mark students in the control group who are adopters (unobserved). Let $Y$ be the student's diagnostic exam score. 
The learning benefit \textbf{E2} can be computed as:
\begin{align}
    \text{\textbf{E2}} &= \frac{1}{N_{c1}} \sum_{i \in c1} Y_i - \frac{1}{N_{c0}} \sum_{i \in c0} Y_i
\end{align}
We can rewrite \textbf{E1} in the same way as well:
\begin{align}
    \textbf{\textbf{E1}} &= \frac{1}{N_1} \br{\sum_{i\in c_1} Y_i + \sum_{i \in n_1} Y_i} -\frac{1}{N_0} \br{\sum_{i\in c_0} Y_i + \sum_{i \in n_0} Y_i}\\
    &= \frac{1}{N_1} \sum_{i\in c_1} Y_i + \frac{1}{N_1} \sum_{i \in n_1} Y_i - \frac{1}{N_0} \sum_{i\in c_0} Y_i - \frac{1}{N_0} \sum_{i \in n_0} Y_i \\
    &= \frac{1}{N_1} \sum_{i\in c_1} Y_i - \frac{1}{N_0} \sum_{i\in c_0} Y_i + \underbrace{\frac{1}{N_1} \sum_{i \in n_1} Y_i - \frac{1}{N_0} \sum_{i \in n_0} Y_i}_{=0} \label{eq:exclusion} \\
     &= \frac{1}{N_1} \sum_{i\in c_1} Y_i - \frac{1}{N_0} \sum_{i\in c_0} Y_i \\
      &= \underbrace{\frac{N_{c1}}{N_1}}_{=P(c)} \frac{1}{N_{c1}} \sum_{i\in c_1} Y_i - \underbrace{\frac{N_{c0}}{N_0}}_{=P(c)} \frac{1}{N_{c0}} \sum_{i\in c_0} Y_i \label{eq:equal_comp} \\
    &= P(c) \br{\frac{1}{N_{c1}} \sum_{i\in c_1} Y_i -  \frac{1}{N_{c0}} \sum_{i\in c_0} Y_i} \\
    &= P(c) \, \text{\textbf{E2}}
\end{align}

As we can see, we can compute \textbf{E2} from \textbf{E1} from this derivation without ever needing to know exactly who the compliers would be in the control group. There are a few key assumptions that make this derivation work.  Eq~\eqref{eq:exclusion} holds from the exclusion principle (Assumption~\thref{ass:exclusion}), which states that the student's exam score is independent of the nudge (advertisement) given their GPT-4 usage status. Second, Eq~\eqref{eq:equal_comp}, the two terms $\frac{N_{c1}}{N_1}$ and $\frac{N_{c0}}{N_0}$ are estimators of the same probability -- the probability of a student being an adopter in the experiment and the control group. As in Assumption~\thref{ass:urn},  we assigned students to the experiment and control group randomly, and therefore the percentage of compliers in the experiment and control group should be identical in expectation, $P(c)$. This term can be estimated as $P(c) = \frac{N_{c1}}{N_1}$.

\section{Conclusion}

Our aim is to understand the impact of introducing access to support of chat-based Generative AI system in an introductory online programming course.  This type of interface is readily available to students and teachers around the world, but there are many important questions about the impact of such tools on engagement and learning. To help tackle such questions, we conducted a large-scale randomized control trial study on an introductory programming class with 5,831 students from 146 countries in which we provided some students with access to a GPT-4 tool for the course. We estimate positive benefits on exam performance for adopters, students who used the tool, but over all students the advertisement led to an average decrease in participation in several class elements. The reason for the decrease in engagement is not yet fully known, and may be influenced by the particular advertisement framing used in sharing access to the tool. Our results suggest there may be promising benefits to using LLMs in introductory courses, but also potential harms for engagement, which may have longer term implications on student success. Our work highlights the need for additional investigations to help understand the potential impact of future large adoption and integration of LLMs into classrooms.

\section*{Acknowledgment}
The research reported in this paper was supported in part by a Stanford HAI Hoffman-Yee grant. 

We thank Tong Mu for project ideation, connections, and discussions. We thank Sierra Wang for her help with the HDI-related data analysis. We thank Rose E Wang, Joy Yueya-He, Yunsung Kim, Lucy Li, David Hall, Yifan Mai, Dora Demszky, Yann HICKE, and Jonathan N. Lee for the discussions. We thank Lama Ahmad and Elizabeth Proehl at OpenAI for giving us a \$15,000 unrestricted grant to support this research. We appreciate their commitment to support global education using LLMs.

\bibliography{refs}

\begin{thebibliography}{}

\bibitem[Angrist, 1990]{angrist1990lifetime}
Angrist, J.~D. (1990).
\newblock Lifetime earnings and the vietnam era draft lottery: evidence from social security administrative records.
\newblock {\em The american economic review}, pages 313--336.

\bibitem[Angrist et~al., 1996]{angrist1996identification}
Angrist, J.~D., Imbens, G.~W., and Rubin, D.~B. (1996).
\newblock Identification of causal effects using instrumental variables.
\newblock {\em Journal of the American statistical Association}, 91(434):444--455.

\bibitem[Athey and Imbens, 2016]{athey2016recursive}
Athey, S. and Imbens, G. (2016).
\newblock Recursive partitioning for heterogeneous causal effects.
\newblock {\em Proceedings of the National Academy of Sciences}, 113(27):7353--7360.

\bibitem[Athey and Imbens, 2017]{athey2017econometrics}
Athey, S. and Imbens, G.~W. (2017).
\newblock The econometrics of randomized experiments.
\newblock In {\em Handbook of economic field experiments}, volume~1, pages 73--140. Elsevier.

\bibitem[Bansak, 2020]{bansak2020generalized}
Bansak, K. (2020).
\newblock {A Generalized Approach to Power Analysis for Local Average Treatment Effects}.
\newblock {\em Statistical Science}, 35(2):254 -- 271.

\bibitem[Bigman et~al., 2021]{bigman2021pearprogram}
Bigman, M., Roy, E., Garcia, J., Suzara, M., Wang, K., and Piech, C. (2021).
\newblock Pearprogram: A more fruitful approach to pair programming.
\newblock In {\em Proceedings of the 52nd ACM Technical Symposium on Computer Science Education}, pages 900--906.

\bibitem[Bloom, 1984]{bloom19842}
Bloom, B.~S. (1984).
\newblock The 2 sigma problem: The search for methods of group instruction as effective as one-to-one tutoring.
\newblock {\em Educational researcher}, 13(6):4--16.

\bibitem[Butgereit and Martinus, 2023]{butgereit2023prof}
Butgereit, L. and Martinus, H. (2023).
\newblock Prof pi: Using whatsapp bots and gpt-4 for tutoring mathematics in underserved areas.
\newblock In {\em International Conference on Innovations and Interdisciplinary Solutions for Underserved Areas}, pages 278--289. Springer.

\bibitem[Card, 1993]{card1993using}
Card, D. (1993).
\newblock Using geographic variation in college proximity to estimate the return to schooling.

\bibitem[Chernozhukov et~al., 2018]{chernozhukov2018double}
Chernozhukov, V., Chetverikov, D., Demirer, M., Duflo, E., Hansen, C., Newey, W., and Robins, J. (2018).
\newblock Double/debiased machine learning for treatment and structural parameters.

\bibitem[Choi et~al., 2023]{choi2023llms}
Choi, J.~H., Garrod, O., Atherton, P., Joyce-Gibbons, A., Mason-Sesay, M., and Bj{\"o}rkegren, D. (2023).
\newblock Are llms useful in the poorest schools? theteacherai in sierra leone.
\newblock {\em arXiv preprint arXiv:2310.02982}.

\bibitem[Cohen and Garcia, 2008]{cohen2008identity}
Cohen, G.~L. and Garcia, J. (2008).
\newblock Identity, belonging, and achievement: A model, interventions, implications.
\newblock {\em Current directions in psychological science}, 17(6):365--369.

\bibitem[Cu and Hochman, 2023]{stanforddaily2023chatgpt}
Cu, M.~A. and Hochman, S. (2023).
\newblock Scores of stanford students used chatgpt on final exams, survey suggests.
\newblock {\em The Stanford Daily}.
\newblock Accessed: 2024-03-20.

\bibitem[Demszky et~al., 2023]{demszky2023can}
Demszky, D., Liu, J., Hill, H.~C., Jurafsky, D., and Piech, C. (2023).
\newblock Can automated feedback improve teachers’ uptake of student ideas? evidence from a randomized controlled trial in a large-scale online course.
\newblock {\em Educational Evaluation and Policy Analysis}, page 01623737231169270.

\bibitem[Duolingo, 2023]{duolingomax2023}
Duolingo (2023).
\newblock Duolingo max uses openai’s gpt-4 for new learning features.
\newblock Accessed: 2024-03-20.

\bibitem[Gemini et~al., 2023]{team2023gemini}
Gemini, Anil, R., Borgeaud, S., Wu, Y., Alayrac, J.-B., Yu, J., Soricut, R., Schalkwyk, J., Dai, A.~M., Hauth, A., et~al. (2023).
\newblock Gemini: a family of highly capable multimodal models.
\newblock {\em arXiv preprint arXiv:2312.11805}.

\bibitem[Gillespie and Hald, 2017]{gillespie2017paradox}
Gillespie, A. and Hald, J. (2017).
\newblock The paradox of helping: Contradictory effects of scaffolding people with aphasia to communicate.
\newblock {\em PLoS One}, 12(8):e0180708.

\bibitem[Greene and DeBacker, 2004]{greene2004gender}
Greene, B.~A. and DeBacker, T.~K. (2004).
\newblock Gender and orientations toward the future: Links to motivation.
\newblock {\em Educational Psychology Review}, 16:91--120.

\bibitem[Hanselman et~al., 2014]{hanselman2014threat}
Hanselman, P., Bruch, S.~K., Gamoran, A., and Borman, G.~D. (2014).
\newblock Threat in context: School moderation of the impact of social identity threat on racial/ethnic achievement gaps.
\newblock {\em Sociology of Education}, 87(2):106--124.

\bibitem[Hansen and Reich, 2015]{hansen2015democratizing}
Hansen, J.~D. and Reich, J. (2015).
\newblock Democratizing education? examining access and usage patterns in massive open online courses.
\newblock {\em Science}, 350(6265):1245--1248.

\bibitem[Imbens and Angrist, 1994]{imbens1994identification}
Imbens, G.~W. and Angrist, J.~D. (1994).
\newblock Identification and estimation of local average treatment effects.
\newblock {\em Econometrica}, 62(2):467--475.

\bibitem[{Ipsos}, 2023]{IpsosGlobalAI2023}
{Ipsos} (2023).
\newblock Global views on a.i. 2023.
\newblock Online.
\newblock A 31-country Global Advisor survey.

\bibitem[Kaplan and Flum, 2012]{kaplan2012identity}
Kaplan, A. and Flum, H. (2012).
\newblock Identity formation in educational settings: A critical focus for education in the 21st century.
\newblock {\em Contemporary Educational Psychology}, 37(3):171--175.

\bibitem[KhanAcademy, 2023]{khanmigo2023}
KhanAcademy (2023).
\newblock Meet khanmigo, khan academy's ai-powered teaching assistant \& tutor.
\newblock Accessed: 2024-03-20.

\bibitem[Kizilcec et~al., 2013]{kizilcec2013deconstructing}
Kizilcec, R.~F., Piech, C., and Schneider, E. (2013).
\newblock Deconstructing disengagement: analyzing learner subpopulations in massive open online courses.
\newblock In {\em Proceedings of the third international conference on learning analytics and knowledge}, pages 170--179.

\bibitem[Kizilcec et~al., 2017]{kizilcec2017closing}
Kizilcec, R.~F., Saltarelli, A.~J., Reich, J., and Cohen, G.~L. (2017).
\newblock Closing global achievement gaps in moocs.
\newblock {\em Science}, 355(6322):251--252.

\bibitem[Kochhar, 2023]{kochhar2023us}
Kochhar, R. (2023).
\newblock Which us workers are more exposed to ai on their jobs?

\bibitem[Kraft, 2020]{kraft2020interpreting}
Kraft, M.~A. (2020).
\newblock Interpreting effect sizes of education interventions.
\newblock {\em Educational researcher}, 49(4):241--253.

\bibitem[Kumar et~al., 2023]{kumar2023math}
Kumar, H., Rothschild, D.~M., Goldstein, D.~G., and Hofman, J.~M. (2023).
\newblock Math education with large language models: Peril or promise?
\newblock {\em Available at SSRN 4641653}.

\bibitem[Lee et~al., 2023]{stanford2023aichatbots}
Lee, V.~R., Spector, C., and Pope, D. (2023).
\newblock What do ai chatbots really mean for students and cheating?
\newblock Accessed: 2024-03-20.

\bibitem[Leite et~al., 2022]{leite2022heterogeneity}
Leite, W.~L., Kuang, H., Shen, Z., Chakraborty, N., Michailidis, G., D'Mello, S., and Xing, W. (2022).
\newblock Heterogeneity of treatment effects of a video recommendation system for algebra.
\newblock In {\em Proceedings of the Ninth ACM Conference on Learning@ Scale}, pages 12--23.

\bibitem[Lens et~al., 2001]{lens2001student}
Lens, W., Simons, J., and Dewitte, S. (2001).
\newblock Student motivation and self-regulation as a function of future time perspective and perceived instrumentality.

\bibitem[Leondari, 2007]{leondari2007future}
Leondari, A. (2007).
\newblock Future time perspective, possible selves, and academic achievement.
\newblock {\em New directions for adult and continuing education}, 114:17--26.

\bibitem[Liu et~al., 2024]{liu2024teaching}
Liu, R., Zenke, C., Liu, C., Holmes, A., Thornton, P., and Malan, D.~J. (2024).
\newblock Teaching cs50 with ai: leveraging generative artificial intelligence in computer science education.
\newblock In {\em Proceedings of the 55th ACM Technical Symposium on Computer Science Education V. 1}, pages 750--756.

\bibitem[Makel and Plucker, 2014]{makel2014facts}
Makel, M.~C. and Plucker, J.~A. (2014).
\newblock Facts are more important than novelty: Replication in the education sciences.
\newblock {\em Educational Researcher}, 43(6):304--316.

\bibitem[Markel et~al., 2023]{markel2023gpteach}
Markel, J.~M., Opferman, S.~G., Landay, J.~A., and Piech, C. (2023).
\newblock Gpteach: Interactive ta training with gpt based students.

\bibitem[Nie et~al., 2023]{nie2023understanding}
Nie, A., Reuel, A.-K., and Brunskill, E. (2023).
\newblock Understanding the impact of reinforcement learning personalization on subgroups of students in math tutoring.
\newblock In {\em International Conference on Artificial Intelligence in Education}, pages 688--694. Springer.

\bibitem[Noy and Zhang, 2023]{noy2023experimental}
Noy, S. and Zhang, W. (2023).
\newblock Experimental evidence on the productivity effects of generative artificial intelligence.
\newblock {\em Available at SSRN 4375283}.

\bibitem[Pardos and Bhandari, 2023]{pardos2023learning}
Pardos, Z.~A. and Bhandari, S. (2023).
\newblock Learning gain differences between chatgpt and human tutor generated algebra hints.
\newblock {\em arXiv preprint arXiv:2302.06871}.

\bibitem[Peteranetz et~al., 2016]{peteranetz2016perceived}
Peteranetz, M.~S., Flanigan, A.~E., Shell, D.~F., and Soh, L.-K. (2016).
\newblock Perceived instrumentality and career aspirations in cs1 courses: Change and relationships with achievement.
\newblock In {\em Proceedings of the 2016 ACM conference on international computing education research}, pages 13--21.

\bibitem[Prihar et~al., 2023]{prihar2023comparing}
Prihar, E., Lee, M., Hopman, M., Kalai, A.~T., Vempala, S., Wang, A., Wickline, G., Murray, A., and Heffernan, N. (2023).
\newblock Comparing different approaches to generating mathematics explanations using large language models.
\newblock In {\em International Conference on Artificial Intelligence in Education}, pages 290--295. Springer.

\bibitem[Rogers et~al., 2014]{rogers2014diffusion}
Rogers, E.~M., Singhal, A., and Quinlan, M.~M. (2014).
\newblock Diffusion of innovations.
\newblock In {\em An integrated approach to communication theory and research}, pages 432--448. Routledge.

\bibitem[Ruan et~al., 2024]{ruan2024reinforcement}
Ruan, S., Nie, A., Steenbergen, W., He, J., Zhang, J., Guo, M., Liu, Y., Dang~Nguyen, K., Wang, C.~Y., Ying, R., et~al. (2024).
\newblock Reinforcement learning tutor better supported lower performers in a math task.
\newblock {\em Machine Learning}, pages 1--26.

\bibitem[Wager, 2020]{wager2020stats}
Wager, S. (2020).
\newblock Stats 361: Causal inference.
\newblock Technical report, Technical report, Stanford University, 2020. URL: https://web. stanford. edu~….

\bibitem[Walton and Cohen, 2007]{walton2007question}
Walton, G.~M. and Cohen, G.~L. (2007).
\newblock A question of belonging: race, social fit, and achievement.
\newblock {\em Journal of personality and social psychology}, 92(1):82.

\bibitem[Woodrow et~al., 2024]{woodrow2024ai}
Woodrow, J., Malik, A., and Piech, C. (2024).
\newblock Ai teaches the art of elegant coding: Timely, fair, and helpful style feedback in a global course.
\newblock In {\em Proceedings of the 55th ACM Technical Symposium on Computer Science Education V. 1}, pages 1442--1448.

\end{thebibliography}

\clearpage
\appendix

\renewcommand\thetable{\thesection.\arabic{table}}
\renewcommand\thefigure{\thesection.\arabic{figure}}
\renewcommand{\theequation}{\thesection.\arabic{equation}}
\setcounter{table}{0}
\setcounter{figure}{0}
\setcounter{equation}{0}

\counterwithout{equation}{section}

\section{Appendices}

\subsection{Additional Student Covariate Interaction Effect}

\begin{figure}[ht]
    \centering
    \begin{subfigure}[b]{0.32\linewidth}
        \includegraphics[width=\linewidth]{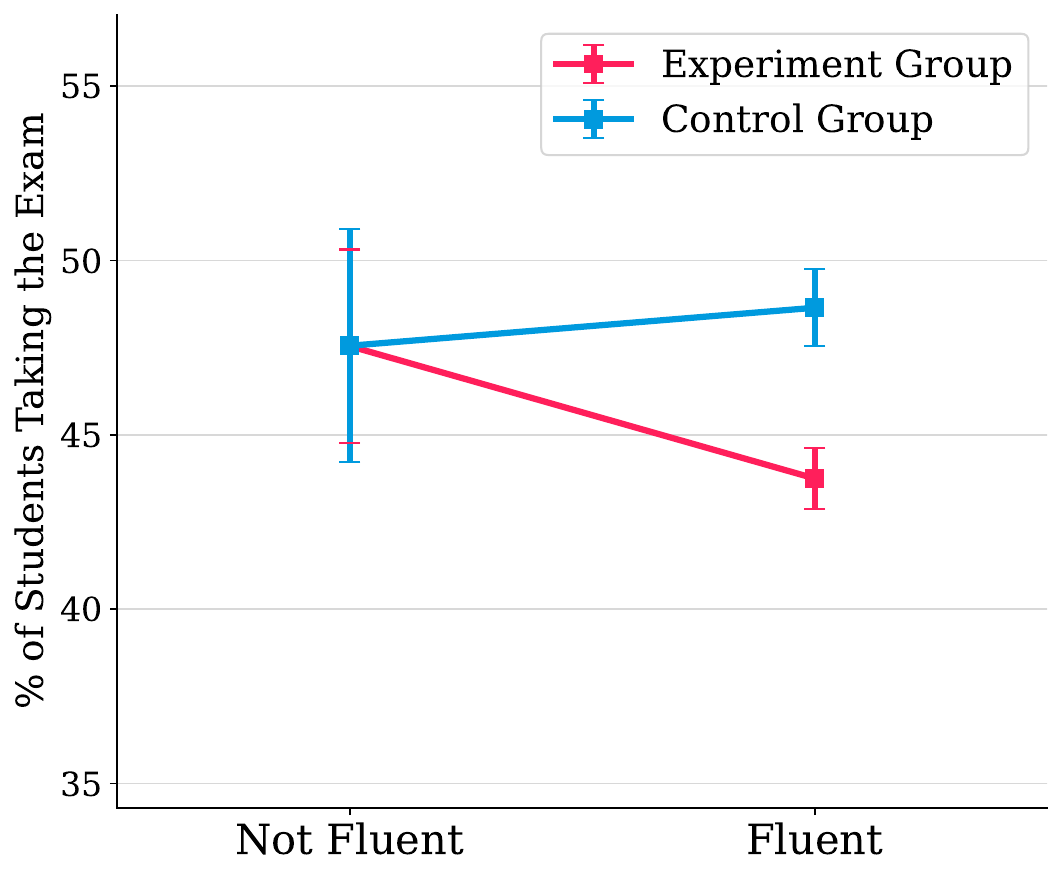}
        \caption{Student English Fluency}
        \label{fig:exam_fluency}
    \end{subfigure}
    \begin{subfigure}[b]{0.32\linewidth}
        \includegraphics[width=\linewidth]{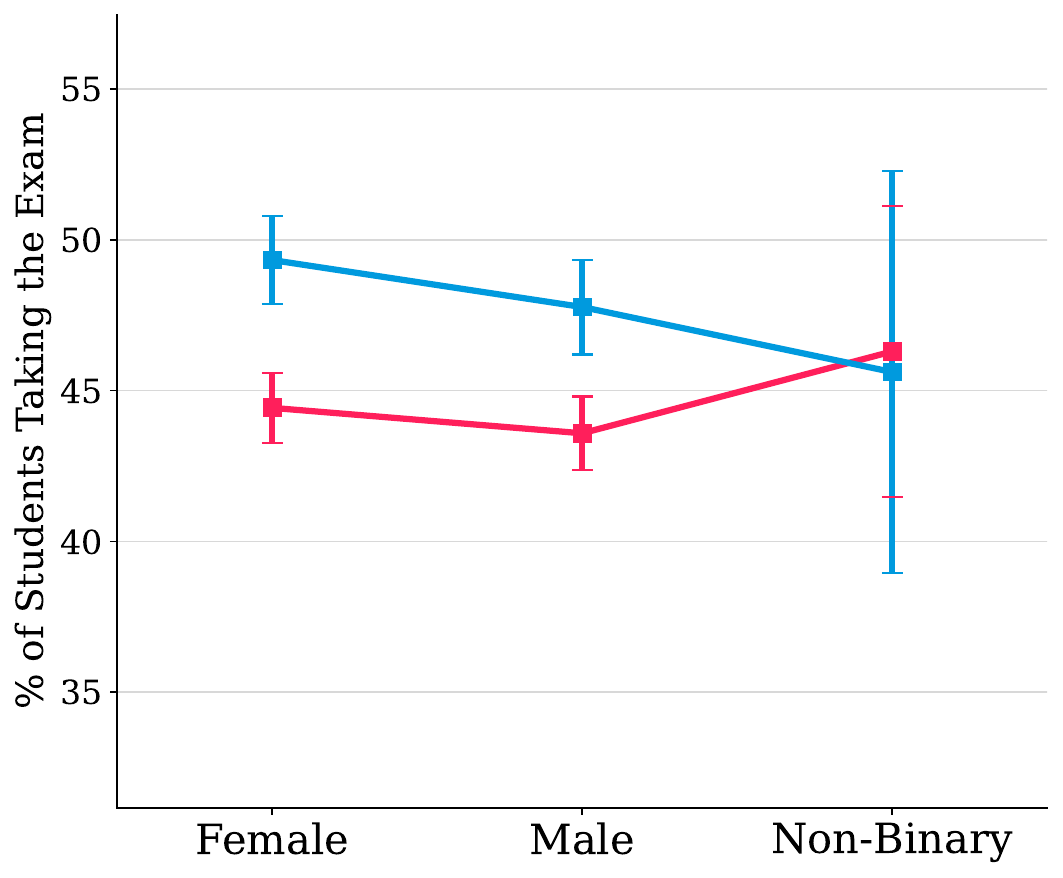}
        \caption{Student Gender}
        \label{fig:exam_gender}
    \end{subfigure}
    \begin{subfigure}[b]{0.32\linewidth}
        \includegraphics[width=\linewidth]{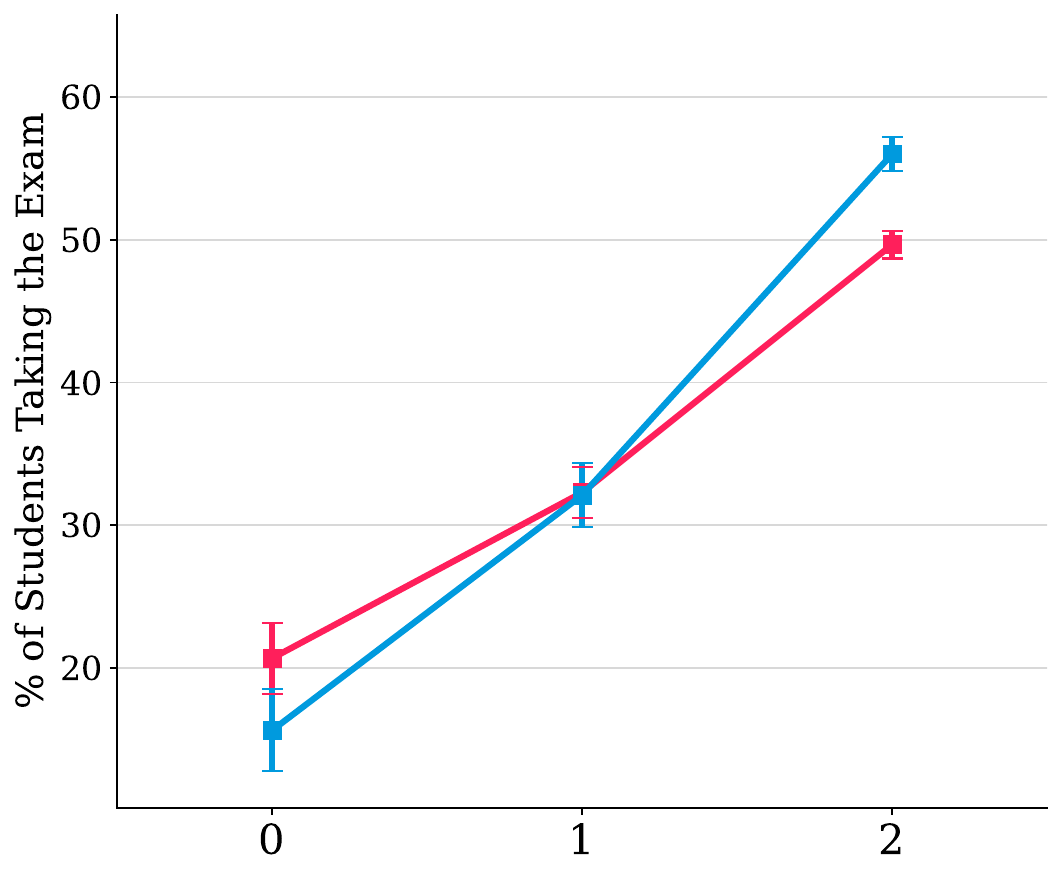}
        \caption{Student Section Attendance}
        \label{fig:exam_sec_att}
    \end{subfigure}
    \caption{We present exploratory analyses for the experiment-control exam participation gap by looking at different student demographics. The vertical line is the standard error.}
    \label{fig:exam_by_covariates_app}
\end{figure}

\begin{figure}[ht]
    \centering
    \includegraphics[width=\textwidth]{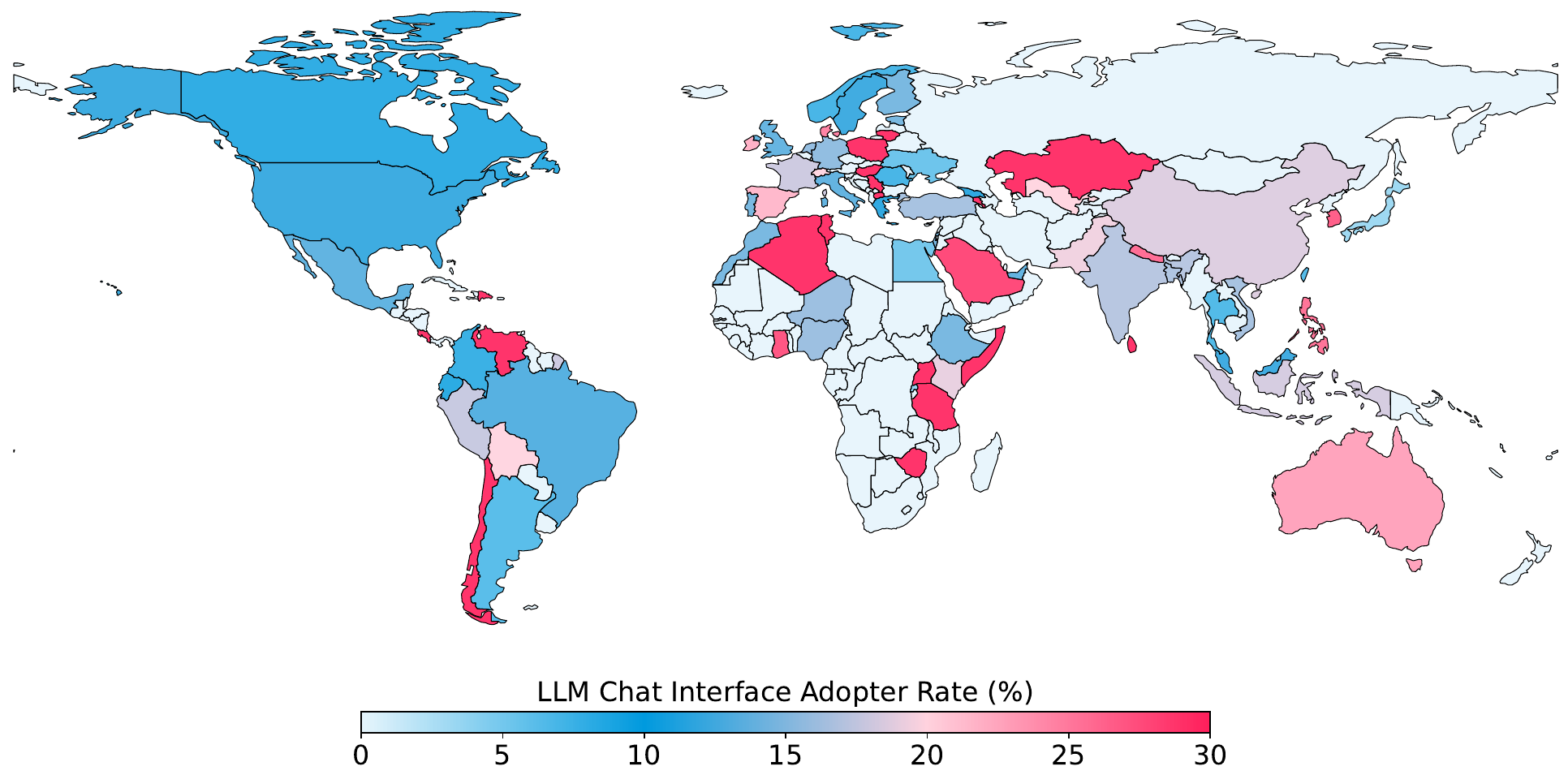}
    \caption{We show the adoption rates of GPT-4, defined as the percentage of students in each country in the experiment who were offered access to GPT-4 and chose to use it. We cap the adoption rate on display at 30\%. We refer to these students as ``adopters''. The overall average adoption rate across all countries is 14.2\%.  The top five countries with the highest rates of adoption, each with at least 10 students enrolled, are Serbia (44.4\%), Hungary (42.9\%), Kazakhstan (40.0\%), Chile (37.5\%), and Poland (35.7\%). Conversely, there are 121 countries with more than 10 students where no students chose to adopt GPT-4. The countries with the lowest nonzero adoption rates include Japan (3.6\%), Rwanda (4.0\%), Egypt (5.6\%), Ukraine (5.9\%), and Argentina (6.9\%).}
    \label{fig:world_map_adoption_rate}
\end{figure}

\begin{figure}[ht]
    \centering
    \includegraphics[width=\textwidth]{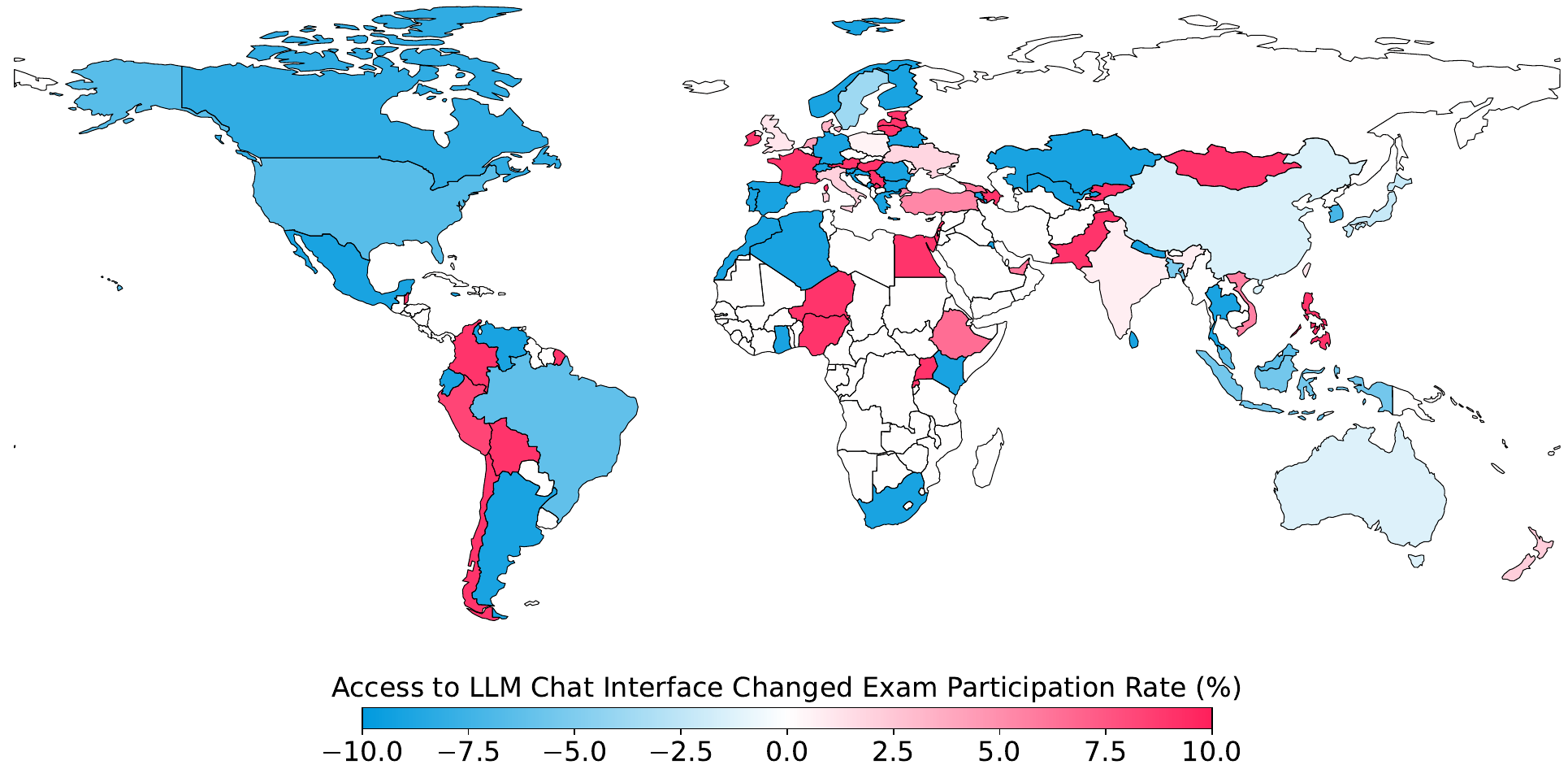}
    \caption{We show the change in exam participation between the experiment and control group by student country. The overall average change in exam participation is -4.4\% across all countries.  The top five countries with the highest positive change, each with at least 20 students enrolled, are Hungary (32.1\%), Pakistan (29.4\%), Egypt (23.6\%), Philippines (16.5\%), and France (13.6\%). The countries with the highest negative change in example participation include Mexico (-22.6\%), Thailand (-22.6\%), Portugal (-23.8\%), Argentina (-25.3\%), and South Africa (-28.7\%). We capped the highest positive or negative change to +10\% and -10\% for the plot.}
    \label{fig:world_map_participation_rate}
\end{figure}

We show additional interactions between student covariates and their exam participation in Figure~\ref{fig:exam_by_covariates_app}. We note that there doesn't seem to be a difference in exam participation for students who are not fluent in English. Nonetheless, as the main instructional materials for this course are in English, most students who decide to apply already have high proficiency in the language, regardless of whether it is the primary language in their home countries. For gender, we notice a similar gap between experiment and control. For section attendance, we are not able to draw meaningful conclusions beyond what we have observed so far.

\subsection{Additional Details on Trial Assignment}

At the beginning of April 24th, 8,762 students enrolled in the class. However, we don't want to offer access to ChatGPT too early because some empirical work in education showed that providing too many hints too early might hurt a student's learning progress~\citep{gillespie2017paradox}. We determine a student to be ``active'' by looking at whether they have completed all Week 1 assignments. We end up with 5,831 students in our randomized control trial. We then sent an email to 3,581 students. 2,778 students (77.6\%) opened the email. 539 students (15.1\%) clicked on the link to our custom ChatGPT interface. We obtained institutional review board (IRB) approval for conducting this experiment.

\subsection{Full Description of Student Covariates}

We report the basic statistics of the student population in our randomized control trial in Table~\ref{tab:demographics}. Here, we provide some additional information about them. Note that we are not reporting these distributions over the entire course's student population. They are only computed on the 5,831 students who were deemed active by week 1 and were included in our randomized control trial. In order to make sure all of the covariates we use are independent of the treatment (access to our custom ChatGPT), we only use either demographic information or a record of the student prior to the start of the experiment (week 3).

\begin{itemize}
    \item \textbf{Application Score} (Mean=48.2, SD=11.4, Median (IQR)=47.2 (41.0-55.0), Max=103.0): This is an aggregate score computed by the course staff to rank student applications to enroll in this class. It is a weighted combination of many factors. We are not able to share what this equation is.
    \item \textbf{Age} (Mean=31.4, SD=10.4, Median (IQR)=29.0 (23.0-37.0), Max=84.0): Student self-reported age.
    \item \textbf{Gender} (``Female'': 51.59\%, ``Male'': 45.58\%, ``Non-Binary/Other/NA'': 2.83\%): Student self-reported gender. We provide five categories in our application.
    \item \textbf{English Fluency} (Mean=13.8, SD=2.5, Median (IQR)=14.0 (12.0-16.0), Max=20.0): This course iteration provided all lectures and materials exclusively in English. Applicants were categorized based on the English fluency demonstrated in their application materials. In future iterations, the course will offer materials in multiple languages to accommodate those who may not be fluent in English.
    \item \textbf{Application Effort} (Mean=4.0, SD=1.3, Median (IQR)=5.0 (3.0-5.0), Max=5.0): The course staff computed a score that captures how much effort an applicant spent on their application. The exact computation is kept confidential. 
    \item \textbf{Coding Score} (Mean=8.2, SD=3.8, Median (IQR)=10.0 (10.0-10.0), Max=10.0): Applicants were required to complete a lesson and then tackle a programming exercise. Their performance on this exercise determined their coding score. In this scoring system, a lower score indicates a lower proficiency in programming, whereas a higher score demonstrates better programming skills.
    \item \textbf{Prior Experience} (Mean=-5.1, SD=4.4, Median (IQR)=-4.0 (-8.0--2.0), Max=0.0): This course is designed for those with little or no prior programming experience. During the admission process, course instructors evaluate applicants' previous programming knowledge to determine if they are overqualified. Applicants detail their prior programming experience in their applications, which is then used to categorize them. A lower, or more negative, score indicates more prior programming experience, which is disadvantageous for their overall evaluation as the course is designed for beginners. Whereas, a higher score indicates less prior programming experience, indicating that the applicant might be better suited for the course.
    \item \textbf{Friend Score} (Mean=4.6, SD=18.3, Median (IQR)=0.0 (0.0-0.0), Max=288.2): Student is asked to tell us if they know other people in the class. We compute a score based on this information. The more people they know, the higher their friend score will be.
    \item \textbf{Section Attendance} (Mean=1.7, SD=0.6, Median (IQR)=2.0 (1.0-2.0), Max=2.0): We calculate how many sections the student has attended prior to receiving access to our custom ChatGPT interface. At most, they could have attended 2 sections.
    \item \textbf{Country HDI} (Mean=0.8, SD=0.1, Median (IQR)=0.9 (0.8-0.9), Max=1.0): Students are asked to self-report their residing country. We map the self-reported country to the United Nations Human Development Index score. This score is a measure of a country's progress on key elements of human development, including health, education, and economic situation\footnote{https://hdr.undp.org/data-center/human-development-index}.
\end{itemize}

\subsection{Diagnostic Exam Details}

A diagnostic exam is administered near the end of the course. All students enrolled in the class received an email notifying them that the exam is available on May 26th, 2023, at 9:43 am EDT. The email was sent to 9,573 students. 7,084 students opened the email (74.4\%), and 1,748 students clicked on the diagnostic exam link in the email (18.4\%). The student will see a welcome message once they open the exam page:

\begin{figure}[htbp]
\begin{tcolorbox}[colback=white!5!white,colframe=black!75!black,breakable,pad at break*=1mm,width=0.9\textwidth,center]
This diagnostic has \textbf{five} questions. Complete each question, to the best of your ability. When you are done, hit the blue Submit button. You can change questions using the numbers in the navbar above. You may go back and forth between questions.

You have 3 hours to complete it from the time you hit start. The diagnostic is designed to only take 50 minutes. Time does not pause if you close the diagnostic and come back to it. For pedagogical purposes, we do allow you to run your programs, but we will not be giving you live feedback as to whether or not your program works.
\end{tcolorbox}
\end{figure}

Questions cover basic concepts of Python knowledge, control flow, arithmetic, and using Python canvas to animate objects. No official score is given to the students. The exam was graded with unit tests that verified each part of the student code, and a rubric system was used to create detailed feedback.

\begin{table}[ht]
\centering
\begin{tabular}{@{}rcccccc@{}}
\toprule
           Diagnostic & Q1 & Q2 & Q3 & Q4 & Q5 & Total \\ \midrule
Rubric Items &  12 & 12 & 14 & 29 & 6 & 73  \\ \bottomrule
\end{tabular}
\end{table}

We use a simple normalization rule to convert a student's diagnostic exam feedback to a score that has a range of 0 to 100. If a student has completed all exam questions and received 0 feedback, they get a score of 100. If a student did not submit a particular question, we count it as if the student received the maximal number of feedback from that question (i.e., if a student misses Q2, we would treat it as if they received 12 feedback). If a student gets $s$ number of feedback in total, their score is $1 - s / 73$. We verified our conversion with the course staff and obtained their approval.

\subsection{Additional Details on Statistics}

In the main result section, we report two $p$-values. One $p$-value is the family-wise error rate (FWER) controlled $p$-value, which we denote as $P$ in the main text. We use Bonferroni correction to control for family-wise error rate. In addition, we report the unadjusted $p$-value per comparison, which we denote as ``unadjusted $P$''. In all the figures, we report the significance level based on the Bonferroni-corrected $P$, which is calculated by multiplying 15 to unadjusted $p$-values. For GPT-4 usage patterns reported in Section~\ref{sec:usage_analysis}, we follow the guideline that the $p$-values for logistic regression analysis do not need to be additionally adjusted. 

For confidence interval, when we use a difference-in-means (DM) estimator for computing $\Delta$ between two groups without missing values (Section~\ref{sec:low_engagement}), we use the standard confidence interval calculation for the DM estimator discussed in \cite{wager2020stats}. When we have to deal with missing data, for both the DM estimator and local average treatment effect (LATE) estimator, we use the bias-corrected and accelerated (BCa) bootstrap interval (Section~\ref{sec:exam_score}). The number of bootstrap samples we used is 1000, and we sampled with replacement.

\subsection{Impute for Missingness (Regression Model)}

All of these features are standardized, which means for feature $X$ with empirical mean $\mu$ and standard deviation $\sigma$, we use $\tilde X = \frac{X - \mu}{\sigma}$. We first discuss how we conduct our model selection (hyperparameter search) and then discuss how we imputed for missing values. We conducted a search over a few model classes: linear regression, Ridge regression, Lasso regression, a 2-layer neural network with 128-dimension hidden size and tanh activation function, and a random forest regressor. All the models are implemented in sklearn. We first split the dataset into a training and a holdout set and used 5-fold cross-validation to select the best model from the training set. We then compute the mean-squared error (MSE) on the holdout test set. Because we do not have access to students who did not participate in the exam, we only train and evaluate our models on students who took the exam (on both the training and holdout set).

\begin{table}[h]
\centering 
\begin{tabular}{@{}lc@{}}
\toprule
Model Name & MSE (Holdout set) \\ \midrule
Linear     & 0.0370            \\
Ridge      & \textbf{0.0355}    \\ 
Lasso     &  0.0356             \\
Neural Net & 0.0391 \\
Random Forest & 0.0397 \\
\bottomrule
\end{tabular}
\caption{\textbf{Model Selection}: We use Ridge regression as our model class. And we use 5-fold cross-validation to choose the best model for imputation in our algorithm.}
\end{table}

We use \textit{cross-fitting} to impute for missing values. This procedure is inspired by works in econometrics and double machine learning~\citep{athey2016recursive,chernozhukov2018double}. Cross-fitting allows us to use part of the dataset to estimate nuisance parameters of the imputation model and use the holdout dataset to estimate the parameter of interest for the causal effect estimation model. We refer readers to Page 22 in \citep{wager2020stats} for a more detailed discussion.

\noindent\makebox[\textwidth][c]{
\begin{minipage}{0.6\textwidth}
\centering
\begin{algorithm}[H]
\SetAlgoLined
\SetKwFunction{split}{Split}
\SetKwFunction{cv}{Cross-Validate}
\SetKwFunction{late}{LATE}
\KwIn{Dataset $\mathcal{D}$, imputation algorithm $\mathcal{A}$, factor $k$}
\KwOut{Estimated Local Average Treatment Effect $\hat \gamma$}
\BlankLine
$\mathcal{I}_1, \mathcal{I}_2, ..., \mathcal{I}_k$ = \split($\mathcal{D}$) \\
\For{$i \leftarrow 1...k$}{
    $\hat D$ = $\emptyset$ \Comment{Missing value imputed dataset} \\ 
    $\hat f$ = \cv($\mathcal{A}, \mathcal{I}_i$) \Comment{Nuisance model} \\
    \For{$x \leftarrow \mathcal{I}_{j \not= i}$}{ 
        $\hat y$ = $\hat f(x)$  \Comment{Imputation}\\
        $\hat D$ = $\hat D \cup \{\hat y\}$  
    }
    $\hat \gamma$ = \late($\hat D$) \Comment{Causal Effect Estimation}
}
 $\hat \gamma$ = $\frac{1}{k} \sum_{i=1}^k \hat \gamma_i$ \\
 \Return{$\hat \gamma$}
 \caption{Cross-fitting based Impute for Missing Values}
 \label{alg:cross-fitting}
\end{algorithm}
\end{minipage}
}

We describe our algorithm above, and we choose $k=2$ for a 2-fold cross-fitting, a standard choice for most settings. The imputation algorithm $\mathcal{A}$ is pre-selected during the model selection phase (the algorithm with the lowest MSE on the holdout is chosen, which, in our case, is the Ridge regression model). 

\subsection{Logistic Regression Analysis of GPT Usage}
In Section~\ref{sec:usage_analysis}, we used a logistic regression model to understand what kind of students are more likely to become adopters of LLMs if we offer to them in a massive online class like ours. Each feature has been standardized. We report the coefficients in Table~\ref{tab:usage_regression}.

\begin{table}[ht]
\centering
\begin{tabular}{lcccccc}
\hline
Variable & Coef. & Std. Err. & $z$ & $P>|z|$ & [0.025 & 0.975] \\ 
\hline
Intercept & -1.8312 & 0.049 & -37.091 & 0.000 & -1.928 & -1.734 \\
Application Score & -0.0238 & 0.063 & -0.376 & 0.707 & -0.148 & 0.100 \\
Gender (Female) & 0.0988 & 0.049 & 2.019 & 0.043 & 0.003 & 0.195 \\
Age & 0.1422 & 0.052 & 2.738 & 0.006 & 0.040 & 0.244 \\
English Fluency & -0.0447 & 0.058 & -0.765 & 0.444 & -0.159 & 0.070 \\
Application Effort & 0.0482 & 0.061 & 0.797 & 0.426 & -0.070 & 0.167 \\
Coding Score & 0.0445 & 0.054 & 0.831 & 0.406 & -0.060 & 0.150 \\
Friend Score & 0.0456 & 0.046 & 0.984 & 0.325 & -0.045 & 0.136 \\
Section Attendance & 0.2126 & 0.055 & 3.840 & 0.000 & 0.104 & 0.321 \\
Country HDI & -0.1856 & 0.052 & -3.561 & 0.000 & -0.288 & -0.083 \\
Prior Experience & 0.0093 & 0.055 & 0.169 & 0.866 & -0.098 & 0.117 \\
\hline
\end{tabular}
\caption{Logistic regression for predicting student GPT usage. Pseudo R-squred is 0.01414. Number of observations is 3581.}
\label{tab:usage_regression}
\end{table}

\subsection{Deployment Implementation Details}
\label{sec:app:deployment}

\paragraph{Student Access to GPT Interface within and outside the course} 

We do not record students' browsing histories and, unfortunately, do not know if they have accessed the publicly available GPT interface provided by OpenAI. However, a survey conducted by a concurrent study on the same course asked if the students had used the ChatGPT interface during the course period. Only 2\% of the students responded that they did.

\paragraph{Using GPT-4 to Cheat on the Diagnostic Exam}

Despite our effort to give a stern warning to the students and let them know that their conversation with the GPT is monitored and visible to course staff, we conducted an analysis to see if the students copied and pasted exam questions into our custom interface. We did not find any evidence that the students used our custom interface to cheat. It is worth pointing out that, unlike a university course, this free online class does not incentivize students to cheat on exams -- the course does not offer a letter grade, and the exam does not provide students a score -- only feedback on how they did on each of the problems. The final course completion certificate only mentions if they had made an attempt on the exam.

\begin{figure}[h]
    \centering
    \includegraphics[scale=0.65]{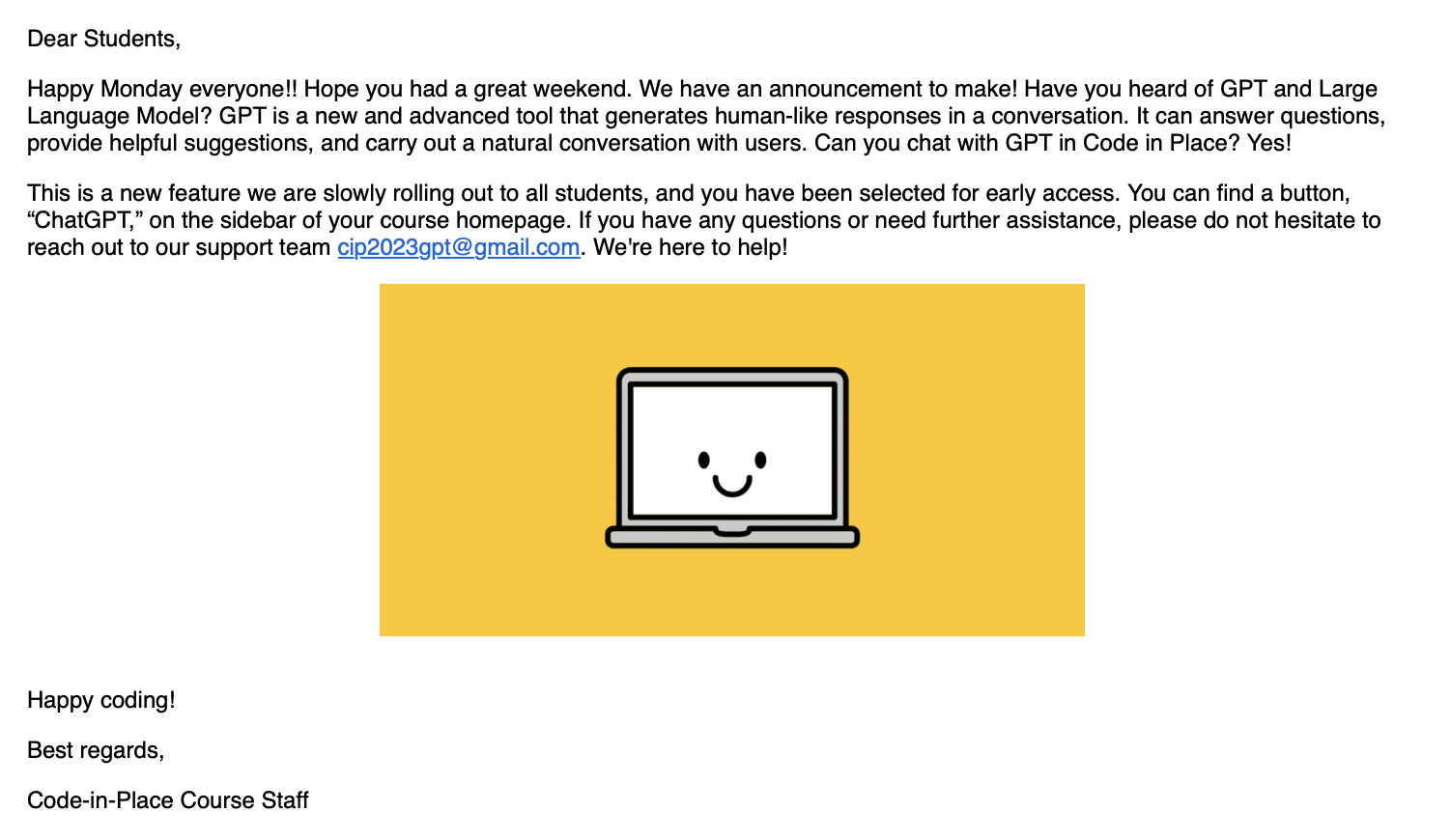}
    \caption{The email sent to alert the students who were granted access to GPT-4.}
    \label{fig:email}
\end{figure}

\begin{figure}
    \centering
    \includegraphics[scale=0.65]{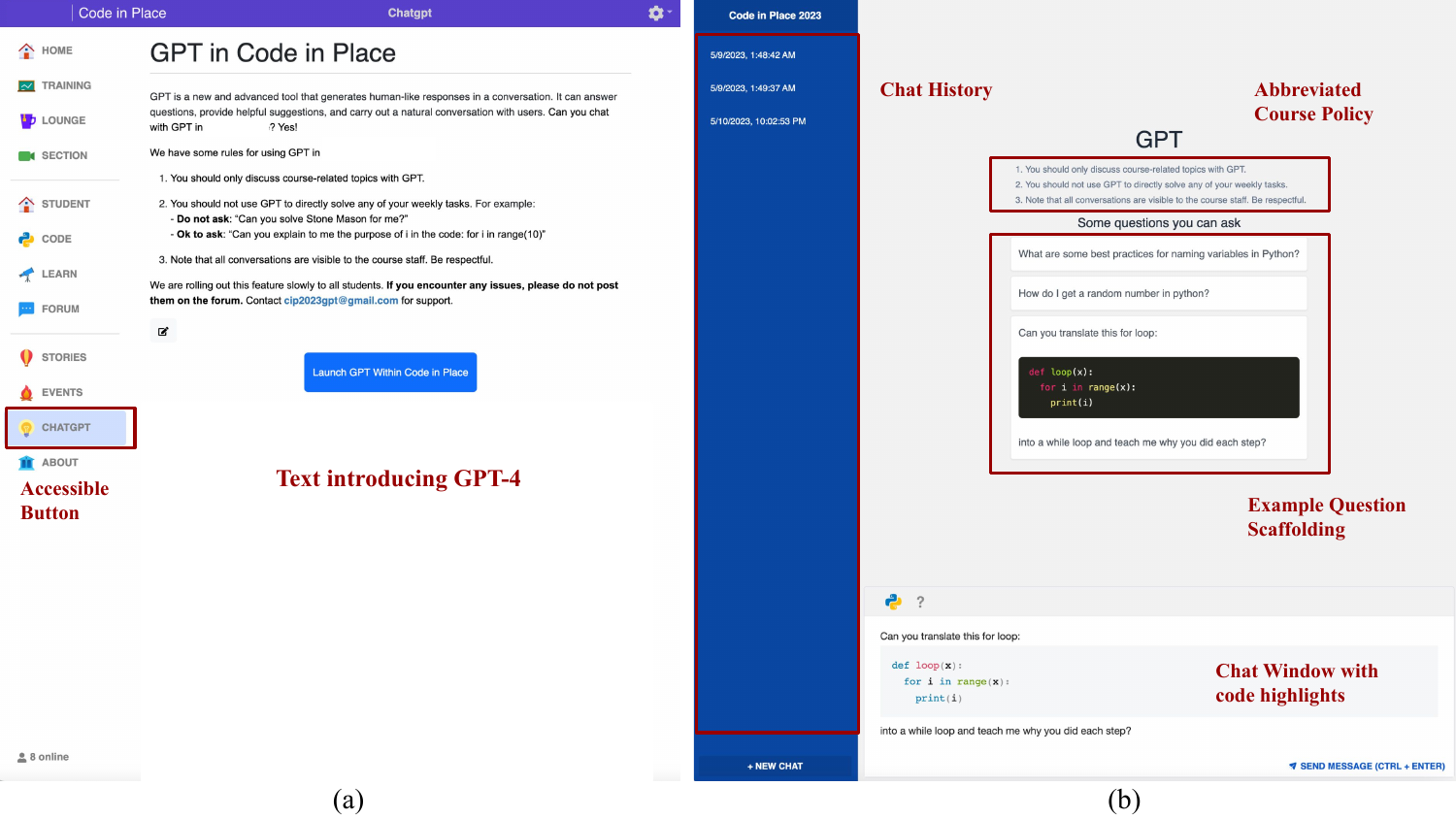}
    \caption{(a) The initial landing page of GPT-4. Students click on the button and have a chance to read through the text introducing GPT-4 before they start using GPT (See Figure~\ref{fig:pos_adv}); (b) The actual chat interface we built. Students can select a pre-generated example question or type in their own question in the chat window. The students who weren't in the treatment group do not see the lightbulb button on their home page.}
    \label{fig:interface}
\end{figure}

\begin{figure}[htbp]
\begin{tcolorbox}[colback=white!5!white,colframe=black!75!black,breakable,pad at break*=1mm,width=0.9\textwidth,center]
  {\Large \textbf{GPT in Code in Place}}

  GPT is a new and advanced tool that generates human-like responses in a conversation. It can answer questions, provide helpful suggestions, and carry out a natural conversation with users. Can you chat with GPT in Code in Place? Yes! 

  We have some rules for using GPT in Code in Place:
  \begin{enumerate}
      \item You should only discuss course-related topics with GPT.
      \item You should not use GPT to directly solve any of your weekly tasks. For example:
      \begin{itemize}
          \item \textbf{Do not ask}: ``Can you solve Stone Mason for me?''
          \item \textbf{Ok to ask}: ``Can you explain to me the purpose of i in the code: for i in range(10)''
      \end{itemize}
  \end{enumerate}
Note that all conversations are visible to the course staff. Be respectful.

\vspace{2mm}
\begin{center}
\begin{tikzpicture}
\node[draw, fill=skyblue, rectangle, rounded corners=5pt, minimum width=2cm, minimum height=1cm, text=black] {Launch GPT Within Code-in-Place};
\end{tikzpicture}
\end{center}

Here is what we think you should know before using a tool like GPT to learn:

\textbf{GPT may provide}:
\begin{itemize}
    \item \textbf{Instant Support}: You can ask questions and receive feedback instantly at any time of day.
    \item \textbf{Context-Based Understanding}: GPT can recall all topics within one conversation, so it can answer you based on what was said before in this interaction. This makes GPT's answers more helpful and personalized to your learning experience in each conversation. 
    \item \textbf{Practice using new tools that may become standard use}: There are more and more tools trying to use GPT to make programming easier. By using GPT in Code in Place, you could get a sense of what this future might look like!
\end{itemize}

However, it could also harm your learning. Here are a few things to consider when using it: Depending too much on GPT can actually make learning how to program more difficult. When GPT does all the work, you might miss an opportunity to develop your understanding of important programming concepts. Another important thing to know is that GPT is sometimes wrong. You will need your strong programming skills to know if the code is correct, and it is your responsibility as a programmer to understand what your code does line by line. Lastly, GPT was trained on data from the entire internet. It may have strong biases and could give harmful or unsettling responses, especially when used outside of the context of the course.

In general, platforms like GPT are powerful but experimental tools. They are not intended as a substitute for professional teaching advice. If you choose to use them, do so at your own risk and with due diligence because they might generate incorrect results.
\end{tcolorbox}
\caption{The paragraphs we display for the student before they enter a custom-built GPT interface.}
\label{fig:pos_adv}
\end{figure}

\end{document}